\RequirePackage{colortbl}
\documentclass{PoS}

\usepackage{amssymb,amsmath,amsbsy}
\usepackage{cite}
\usepackage[utf8x]{inputenc}
\usepackage{float}
\usepackage{graphicx}
\usepackage{psfrag,fancyhdr,epsfig}

\newcommand{\la}{\lambda_1}
\newcommand{\lb}{\lambda_2}
\newcommand{\lc}{\lambda_3}
\newcommand{\g}{\,\mathrm{GeV}}
\newcommand{\be}{\begin{equation}}
\newcommand{\ee}{\end{equation}}

\newcommand{\Vtree}{V^{(0)}}
\newcommand{\Vone}{V^{(1)}}
\newcommand{\VT}{V^{T}}
\newcommand{\f}{\varphi}

\newcommand{\GeV}{\,{\rm GeV}}
\newcommand{\TeV}{\,{\rm TeV}}

\usepackage{physics}
\title{Scale-Invariant Model for Gravitational Waves and Dark Matter}
\ShortTitle{Scale-Invariant Model for GW and DM}

\author{\speaker{Alexandros Karam}\\%\thanks{A footnote may follow.}\\
        Laboratory of High Energy and Computational Physics, \\
        National Institute of Chemical Physics and Biophysics, R\"avala pst. 10, 10143 Tallinn, Estonia;\\
        E-mail: \email{alexandros.karam@kbfi.ee}}

\author{Maciej Kierkla and Bogumi\l a~\'Swie\.zewska\\%\thanks{A footnote may follow.}\\
        Faculty of Physics, University of Warsaw, Pasteura~5, 02-093 Warsaw, Poland;\\
        E-mail: \email{maciej.kierkla@fuw.edu.pl}; \email{bogumila.swiezewska@fuw.edu.pl}}

\abstract{The present contribution summarises the results recently published in Ref.~\cite{Kierkla:2022odc}. We have conducted a revised analysis of the first-order phase transition that is associated with symmetry breaking in a classically scale-invariant model that has been extended with a new $SU(2)$ gauge group. By incorporating recent developments in the understanding of supercooled phase transitions, we were able to calculate all of its features and significantly limit the parameter space. We were also able to predict the gravitational wave spectra generated during this phase transition and found that this model is well-testable with LISA. Additionally, we have made predictions regarding the relic dark matter abundance. Our predictions are consistent with observations but only within a narrow part of the parameter space. We have placed significant constraints on the supercool dark matter scenario by improving the description of percolation and reheating after the phase transition, as well as including the running of couplings. Finally, we have also analyzed the renormalization-scale dependence of our results.}

\FullConference{%
  Corfu Summer Institute 2022 ``School and Workshops on Elementary Particle Physics and Gravity",\\
  28 August - 1 October 2022,\\
  Corfu, Greece}

\begin{document}
\maketitle

%%%%%%%%%%%%%%%%%%%%%%%%%%%%%%%%%%%%%%%%%%%%%%%
%%%%%%%%%%%%%%%%%%%%%%%%%%%%%%%%%%%%%%%%%%%%%%%
\section{Introduction}
%%%%%%%%%%%%%%%%%%%%%%%%%%%%%%%%%%%%%%%%%%%%%%%
%%%%%%%%%%%%%%%%%%%%%%%%%%%%%%%%%%%%%%%%%%%%%%%

Considering the recent direct detection of gravitational waves (GW) by the LIGO and Virgo Collaborations~\cite{Abbott:2016-2, Abbott:2016, Abbott:2017, Abbott:2017-2, LIGOScientific:2017ycc, LIGOScientific:2017vox}, as well as the upcoming Laser Interferometer Space Antenna (LISA)~\cite{Bartolo:2016ami, Caprini:2019pxz, Gowling:2021gcy, LISACosWG:2022jok, Boileau:2022ter, Gowling:2022pzb} and other future and ongoing experiments~\cite{Badurina:2019hst, Graham:2016plp, 
 Graham:2017pmn, AEDGE:2019nxb, Punturo:2010zz, Hild:2010id, Harry:2010zz, VIRGO:2014yos, LIGOScientific:2014pky, LIGOScientific:2019lzm}, it is reasonable to explore ways to utilize GW to investigate fundamental physics. One promising method is to search for evidence of a first-order phase transition (PT) in the early Universe through the primordial gravitational wave background~\cite{Caprini:2015, Caprini:2019pxz, LISACosWG:2022jok, Gowling:2021gcy, Boileau:2022ter, Gowling:2022pzb}. This signal is expected to be present at frequencies within LISA's sensitivity range if the transition occurred around temperatures similar to those of the electroweak PT, $T\sim 100\ \g$. However, in many models, the signal is not strong enough to be detected. In contrast, the class of models with classical scale invariance%, which includes models with strong dynamics, extra dimensions, and perturbative classically scale-invariant potentials
 ~\cite{Hambye:2013, Jaeckel:2016, Hashino:2016, Jinno:2016, Marzola:2017, Ghorbani:2017lyk, Baldes:2018, Prokopec:2018, Marzo:2018, Mohamadnejad:2019vzg, Ghoshal:2020vud, Kang:2020jeg, Mohamadnejad:2021tke, Dasgupta:2022isg} typically predicts a strong gravitational wave signal within LISA's reach due to a logarithmic potential that enables significant supercooling and latent heat release during the transition.

 Within the wide variety of classically conformal models, those incorporating an additional gauge group are particularly promising due to their high level of predictability. The conformal Standard Model (SM) can be extended in a minimal manner with the addition of either an extra $U(1)$~\cite{Hempfling:1996, Sher:1996ib, Chang:2007, Iso:2009, Iso:2012jn, Khoze:2013-1, Khoze:2013-2, Khoze:2013-3, Hashimoto:2013hta, Hashimoto:2014ela, Benic:2014xho, Khoze:2014, Benic:2014aga, Okada:2014nea, Guo:2015, Humbert:2015epa, Oda:2015gna, Humbert:2015yva, Plascencia:2015, Haba:2015lka, Das:2015nwk, Haba:2015nwl, Wang:2015sxe, Jinno:2016, Das:2016zue, Oda:2017kwl, Hambye:2018, Loebbert:2018, Marzo:2018, YaserAyazi:2019caf, Kim:2019ogz, Mohamadnejad:2019vzg, Kang:2020jeg, Gialamas:2021enw, Barman:2021lot, Barman:2203, Mohamadnejad:2021tke, Dasgupta:2022isg} or $SU(2)$~\cite{Hambye:2013, Carone:2013, Khoze:2014, Pelaggi:2014wba, Karam:2015, Plascencia:2016, Chataignier:2018, Hambye:2018, Baldes:2018, Prokopec:2018, Marfatia:2020} symmetry, and there are other possibilities such as models featuring an extended scalar sector~\cite{Sher:1996ib, Meissner:2006, Foot:2007s, Foot:2007-3, Foot:2007, Foot:2010av, AlexanderNunneley:2010, Foot:2010et, Lee:2012, Farzinnia:2013, Gabrielli:2013, Steele:2013fka, Guo:2014, Salvio:2014soa, Khoze:2014, Davoudiasl:2014, Farzinnia:2014xia, Lindner:2014oea, Kang:2014cia, Kannike:2015apa, Endo:2015ifa, Kang:2015aqa, Endo:2015nba, Ahriche:2015loa, Wang:2015cda, Ghorbani:2015xvz, Farzinnia:2015fka, Helmboldt:2016, Ahriche:2016cio, Ahriche:2016ixu, Wu:2016jdo, Marzola:2017, Ghorbani:2017lyk, YaserAyazi:2018lrv, Oda:2018zth, Brdar:2018, Brdar:2018num, Mohamadnejad:2019wqb, Kannike:2019upf, Jung:2019dog, Brdar:2019qut, Braathen:2020vwo, Kannike:2020ppf, Kubo:2020fdd, Ahriche:2021frb, Soualah:2021xbn}, larger gauge groups, extra fermions, or more intricate architectures~\cite{Dias:2005jk, Holthausen:2009uc, Heikinheimo:2013, Dermisek:2013, Holthausen:2013ota, Kubo:2014ida, Altmannshofer:2014, Antipin:2014qva, Giudice:2014tma, Ametani:2015jla, Carone:2015jra, Kubo:2015joa, Latosinski:2015pba, Haba:2015qbz, Karam:2016, Kubo:2016, Ishida:2019gri, Dias:2020ryz, Aoki:2020mlo, Dias:2022hbu}. The focus of our current work is on the first-order PT in a classically scale-invariant model that includes an additional $SU(2)_X$ gauge symmetry and a scalar that transforms as a doublet under this group while remaining a singlet of the SM. In addition to exhibiting a strong first-order phase transition, this model also provides a candidate for dark matter particles that are stabilized by a residual symmetry that persists after the $SU(2)_X$ symmetry is broken~\cite{Gross:2015, Hambye:2008}.

Although the possibility of detecting GW from a PT and exploring events that occurred in the early Universe is exciting, the imprecise nature of theoretical predictions is discouraging~\cite{Croon:2020, Athron:2022}. The dependence on the renormalisation scale is one of the main sources of uncertainty in these predictions. Classically scale-invariant models, owing to the logarithmic nature of their potential, span a broad range of energies and therefore are particularly susceptible to issues related to scale dependence. In this work~\cite{Kierkla:2022odc}:
\begin{enumerate}
    \item We present updated predictions of the stochastic GW background in the classically scale-invariant model with $SU(2)_X$ symmetry, incorporating recent advances in understanding supercooled PTs~\cite{Ellis:2018, Ellis:2019, Lewicki:2019, Ellis:2020-2, Ellis:2020}. %The nucleation condition is accurately formulated and we investigate whether the PT successfully ends with percolation, which is not guaranteed in models with strong supercooling. 
    Our study is the first to include the condition for percolation in the SU(2)$_X$ model, and we show that it significantly affects the parameter space. 
    %The GW spectra are evaluated using updated simulations~\cite{Caprini:2019pxz, Lewicki:2020, Lewicki:2022pdb}, and we determine the dominant source (sound waves vs bubble collisions) using recent developments~\cite{Ellis:2019, Ellis:2020, Hoche:2020ysm, Gouttenoire:2021kjv}.

    \item We pay close attention to the renormalisation-scale dependence of the results. To minimise this dependence, we use a renormalisation-group improved effective potential and perform an expansion in powers of couplings consistent with the conditions from conformal symmetry breaking and the radiative nature of the transition. %We also perform separate scans of the parameter space at different fixed renormalisation scales to study the dependence of the results on the arbitrary scale.

    \item We investigate the DM phenomenology in light of the updated understanding of the PT.
\end{enumerate}

%%%%%%%%%%%%%%%%%%%%%%%%%%%%%%%%%%%%%%%%%%%%%%%
%%%%%%%%%%%%%%%%%%%%%%%%%%%%%%%%%%%%%%%%%%%%%%%
\section{The model}
%%%%%%%%%%%%%%%%%%%%%%%%%%%%%%%%%%%%%%%%%%%%%%%
%%%%%%%%%%%%%%%%%%%%%%%%%%%%%%%%%%%%%%%%%%%%%%%
% \be
% \Vtree(\Phi, \Psi)=\la \left(\Phi^{\dagger}\Phi\right)^2 + \lb \left(\Phi^{\dagger}\Phi\right) \left(\Psi^{\dagger}\Psi \right)+ \lc \left(\Psi^{\dagger}\Psi\right)^2,%\notag
% \ee

% \be
% \Phi^{\dagger}\Phi=\frac{1}{2} h^2, \quad \Psi^{\dagger}\Psi=\frac{1}{2} \f^2.\notag
% \ee

In this work~\cite{Kierkla:2022odc}, we analyse the classically scale-invariant SM extended by a dark $SU(2)_X$ gauge group. The new fields of the model are:
\begin{itemize}
    \item the scalar doublet $\Phi$ of $SU(2)_X$,
    \item the three dark gauge bosons $X$ of $SU(2)_X$.
\end{itemize}
The Higgs $H$ and new scalar $\Phi$ doublets can be written as 
\[ 
H = \frac{1}{\sqrt{2}}\left(\begin{array}{c}
0\\
h
\end{array}\right), \quad  \Phi = \frac{1}{\sqrt{2}}\left(\begin{array}{c}
0\\
\f
\end{array}\right)\,.
\]
In terms of $h$ and $\f$, the one-loop effective potential can be written as
\be
V(h,\f)=\Vtree(h,\f) + \Vone(h,\f),\label{eq:one-loop-potential}
\ee
where the tree-level part is
\be
\Vtree(h,\f)=\frac{1}{4}\left(\la h^4 + \lb h^2 \f^2 + \lc \f^4\right)\,,
\label{eq:Vtree}
\ee
with $\lb$ being the portal coupling that connects the visible and dark sectors. The one-loop correction is given by 
\be
\Vone(h,\f)=\frac{1}{64 \pi^2}\sum_{a}n_a M_a^4(h,\f)\left(\log\frac{M_a^2(h,\f)}{\mu^2}-C_a\right), \label{eq:one-loop}
\ee
where
\be
n_a=(-1)^{2s_a} Q_a N_a (2s_a+1),\notag%\label{eq:dof}
\ee
and the sum runs over all particle species. With $M_a(h,\f)$ we denote the field-dependent mass of a particle, $n_a$ denotes the number of degrees of freedom associated with each species and $C_a=\frac{5}{6}$ for vector bosons and $C_a=\frac{3}{2}$ for other particles. Furthermore, $Q_a=1$ for uncharged particles, and $Q_a=2$ for charged particles, $N_a=1,\,3$ for uncoloured and coloured particles, respectively.

% \be
%  M^2(h,\f)=\left(\begin{array}{cc}
% 3\la h^2 + \frac{\lb}{2}\f^2		 & \lb h \f\\
% \lb h \f 			& 3 \lc \f^2 + \frac{\lb}{2} h^2\\
%  \end{array}
%  \right), \label{eq:tree-level-mass-matrix}
%  \ee

%  \begin{align}
% M^2_{\pm}(h,\f)& = \frac{1}{2}\left(3 \lambda_1 + \frac{\lambda_2}{2}\right)h^2 + \frac{1}{2}\left(\frac{\lambda_2}{2} + 3\lambda_3\right)\varphi^2  \notag\\[2pt]
% & \ \ \ \pm\frac{1}{2}\sqrt{\left[\left(3\lambda_1-\frac{\lambda_2}{2}\right)h^2-\left(3\lambda_3-\frac{\lambda_2}{2}\right)\varphi^2\right]^2+4\lambda_2^2h^2\varphi^2}.\label{eq:M-plus-minus}
% \end{align}

% \begin{align}
%     M_G^2(h,\f)&=M_{G^{\pm}}^2(h,\f)=\la h^2 + \frac{1}{2}\lb \f^2,\\
%     M_{{G_X}}^2(h,\f)&=M_{G_X^{\pm}}^2(h,\f)=\lc \f^2 + \frac{1}{2}\lb h^2.\label{eq:X-Goldstone-masses}
% \end{align}

% \be
% M_{W^{\pm}}(h,\f)=\frac{1}{2}g_2 h,\quad M_Z(h,\f)=\frac{1}{2} \sqrt{g_2^2+g_Y^2} h,\quad M_X(h,\f)=\frac{1}{2}g_X \f,\quad M_t(h,\f)=\frac{1}{\sqrt{2}} y_t h,
% \ee

Regarding symmetry breaking, the stationary point equations divided by the VEVs, $v=\langle h \rangle$, $w=\langle\f\rangle$, read
\begin{align}
\frac{1}{v^3}\frac{\partial V}{\partial h}=\la&  +\frac{1}{2} \lb \left(\frac{w}{v}\right)^2 +\frac{1}{v^3}\left.\frac{\partial \Vone}{\partial h}\right|_{h=v, \f=w}=0,\label{eq:min1}\\
\frac{1}{w^3}\frac{\partial V}{\partial \f}=\lc& +\frac{1}{2} \lb \left(\frac{v}{w}\right)^2 +\frac{1}{w^3}\left.\frac{\partial \Vone}{\partial \f}\right|_{h=v, \f=w}=0.\label{eq:min2}
\end{align}
Typically, $v_\f/v_h\gg 10$, therefore the $\lb \left( v_h / v_\f \right)^2$ term can be neglected. Then, the second equation becomes
\be
\lc= - \frac{9}{256\pi^2}g_X^4\left[2\log\left(\frac{g_X}{2}\frac{w}{\mu}\right) -\frac{1}{3}\right].\label{eq:min-phi}
\ee
The first equation reads
% \be
% \la +\frac{1}{2} \lb \left(\frac{w}{v}\right)^2=0.\label{eq:tree-level-min}
% \ee
\be
\la +\frac{1}{2} \lb \left(\frac{w}{v}\right)^2 +\frac{1}{16\pi^2}\sum_{W^{\pm},Z,t}n_a \frac{M_a^4(h,\f)}{v^4}\left(\log\frac{M_a^2(h,\f)}{\mu^2}-C_a+\frac{1}{2}\right)=0. \label{eq:condition-lambda2}
\ee
The above indicates that the symmetry breaking in the $\f$ direction follows the Coleman-Weinberg mechanism, while the symmetry breaking in the direction of $h$ is similar to that of the SM, as the "tree-level mass term" is generated by the portal coupling.

The physical mass corresponds to a pole of the propagator, i.e.\ is evaluated away from $p^2=0$, and is given by
\be
M^2_{\mathrm{pole}}=m^2_{\textrm{tree-level}}+\Re[\Sigma(p^2=M^2_{\mathrm{pole}})].\label{eq:pole-mass}
\ee
Including loop corrections from self energies which introduce momentum dependence, we have
\be
 M^2(p)=\left(\begin{array}{cc}
3\la v^2 + \frac{\lb}{2}w^2		 & \lb v w\\
\lb v w 			& 3 \lc w^2 + \frac{\lb}{2} v^2\\
 \end{array}
 \right)
 +
\left(\begin{array}{cc}
\Sigma_{hh}(p)		 & \Sigma_{h\f}(p)\\
\Sigma_{h\f}(p)			& \Sigma_{\f\f}(p)\\
 \end{array}
 \right).
\ee
By diagonalising the mass matrix we obtain the mass eigenvalues
\begin{align}
M^2_{\pm}(p^2)& = \frac{1}{2}\Biggl\{\left(3 \lambda_1 + \frac{\lambda_2}{2}\right)v^2 + \frac{1}{2}\left(\frac{\lambda_2}{2} + 3\lambda_3\right)w^2 +\Sigma_{hh}(p^2)+\Sigma_{\f\f}(p^2) \notag\\[2pt]
& \ \ \ \pm\sqrt{\left[\left(3\lambda_1-\frac{\lambda_2}{2}\right)v^2-\left(3\lambda_3-\frac{\lambda_2}{2}\right)w^2+\Sigma_{hh}(p^2)-\Sigma_{\f\f}(p^2)\right]^2+4\lambda_2^2v^2w^2}\,\Biggr\}.\label{eq:momentum-dependent-masses}
\end{align}
Neglecting terms suppressed by a product of a small coupling, $\lb$ or $\lc$ and the Higgs VEV, we can approximately determine which of the mass eigenvalues corresponds to the Higgs particle. We find
\begin{align}
    &M_{+}^{2}(h,\f) =3 \lc \f^2+ \Sigma_{\f\f}(p^2), \\     &M_{-}^{2}(h,\f) = 3\la h^2 +\frac{1}{2}\lb \f^2+\Sigma_{hh}(p^2)\,.
\end{align}
for $3\la h^2-3\lc \f^2+\frac{1}{2}\lb \f^2+\Sigma_{hh}(p^2)-\Sigma_{\f\f}(p^2)<0$. For the opposite sign, $M_{+}$ and $M_{-}$ are interchanged. Then, to obtain the momentum-corrected masses we solve the gap equations
\begin{align}
    M_H^2&=M_{\mp}^2(p^2=M_H^2),\label{eq:gap-Higgs}\\
    M_S^2&=M_{\pm}^2(p^2=M_S^2).\label{eq:gap-scalar}
\end{align}
We identify the first one with the Higgs $M_H=125\g$, while the other gives the mass of the new scalar $S$. Finally, the mass eigenstates are obtained from the gauge eigenstates by a rotation matrix as
\be
\left(\begin{array}{c}\phi_-\\ \phi_+\end{array}\right)=
\left(\begin{array}{rl}
\cos\theta & \sin\theta\\
-\sin\theta & \cos\theta
\end{array}\right)\left(\begin{array}{c}h \\ \f\end{array}\right)\,, \qquad -\frac{\pi}{2} < \theta < \frac{\pi}{2}\,.
\label{eq:mixing}
\ee

% \begin{equation}
% \xi_H=\left\{\begin{array}{rrl}
%     \cos\theta & \textrm{ for }& M_H\leqslant M_S  \\
%     -\sin\theta & \textrm{ for }& M_H>M_S
% \end{array}\right.,\quad
% \xi_S=\left\{\begin{array}{lrl}
%     -\sin\theta & \textrm{ for }& M_H\leqslant M_S  \\
%     \phantom{-}\cos\theta &\textrm{ for }& M_H>M_S
% \end{array}\right..
% \label{eq:xi}
% \end{equation}

% \be
% g_X(M_Z)\leqslant 1.15, \label{eq:pert-gx}
% \ee

In order to scan the parameter space, we employ the following numerical procedure:
\begin{enumerate}
    \item We choose the values of the input parameters, $M_X$ and $g_X$. We assume the tree-level relation for the $X$ mass $M_X=\frac{1}{2}g_X v_\f$ so we can compute the value of the $\f$ VEV, $v_\f$. %We assume that $M_X$ corresponds to the physical mass of the $X$ bosons. 
    The values of $g_X$ and $v_\f$ are treated as evaluated at the scale $\mu=M_X$.
    
    \item We use the minimisation condition along the $\f$ direction, evaluated at $\mu=M_X$ to evaluate $\lambda_3$. This gives us a simple relation
    \[
        \lambda_3=\frac{3}{256\pi^2}g_X^4.
    \]
    
    \item The $g_X$ and $\lambda_3$ couplings are evolved using their RGEs and evaluated at %the reference scale 
    $\mu=M_Z$.
    
    \item If $g_X(M_Z)\leqslant 1.15$ the RG-improved potential is well-behaved throughout the scales considered.
    
    \item The value of $\lambda_2$ as a function of $\lambda_1(\mu=M_Z)$ %(at $\mu=M_Z$) 
    is obtained from the first minimisation condition.
   
    \item The value of $\lambda_1$ is computed from the requirement that the physical Higgs mass is equal to $125\ \rm{GeV}$, using the first gap equation. The evaluation is performed at $\mu=M_Z$, therefore the vacuum expectation value of $\f$ at $\mu=M_Z$ is needed. It is found using the second minimisation condition evaluated at $\mu=M_Z$. 
    
    \item The mass of $S$ is computed by solving iteratively the second gap equation.
    
    \item The mixing between the scalars is evaluated by demanding that the off-diagonal terms of the mass matrix evaluated at $p^2=0$ and in the mass-eigenbasis are zero.
\end{enumerate}

%
% \begin{figure}[h!t]
% \center
% \includegraphics[width=.45\textwidth]{scan-lambda2.pdf}\hspace{20pt}
% \includegraphics[width=.45\textwidth]{scan-lambda3.pdf}
% \caption{Values of the scalar couplings $\lb$ (left panel) and $\lc$ (right panel) evaluated at the electroweak scale. Gray shaded regions are excluded, from left to right: no electroweak minimum with correct mass and VEV of the Higgs exists, perturbativity of $g_X$ (see eq.~\eqref{eq:pert-gx}).\label{fig:scalar-couplings}}
% \end{figure}
%

%
\begin{figure}[ht]
\center
\includegraphics[width=.45\textwidth]{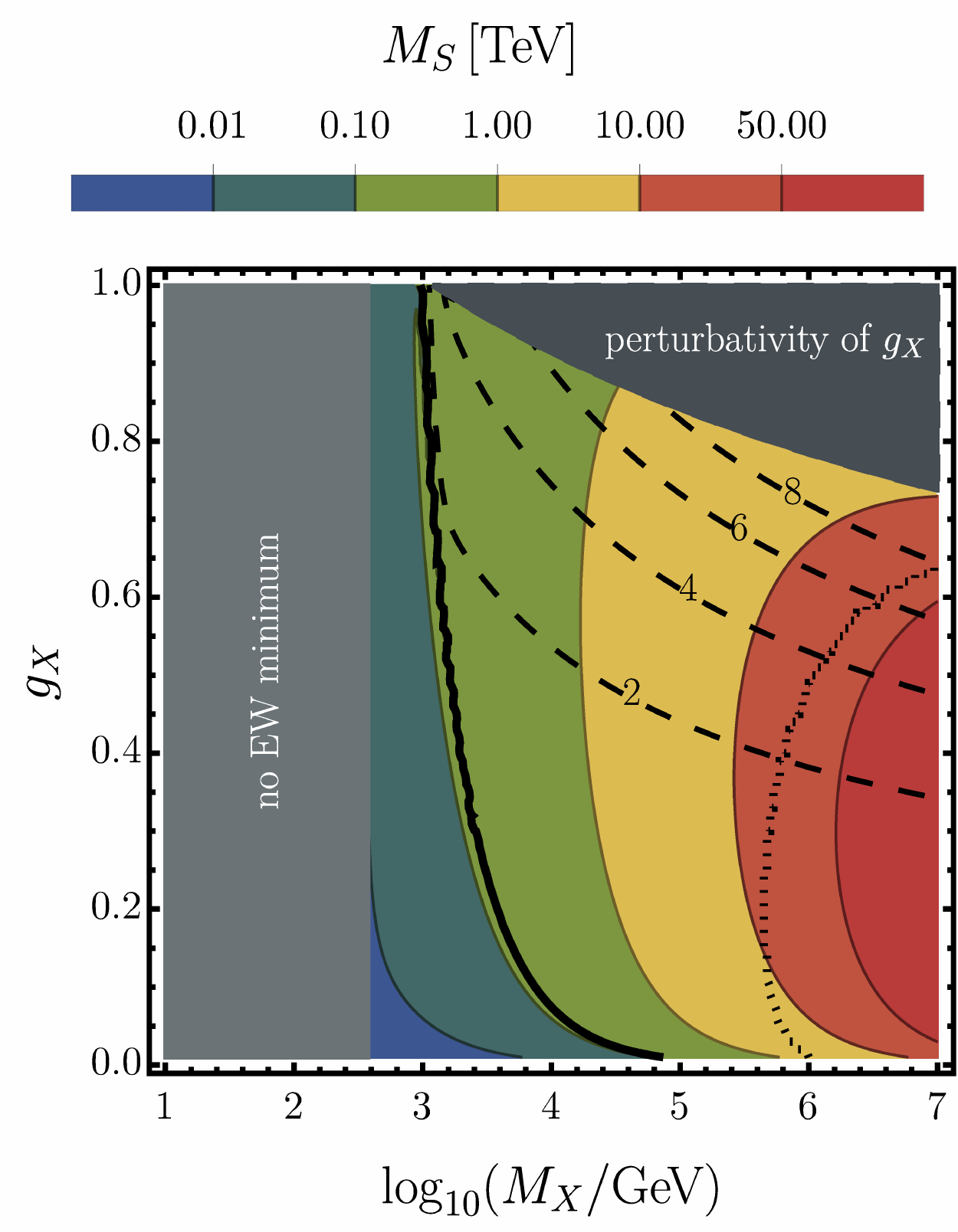}\hspace{20pt}
\includegraphics[width=.45\textwidth]{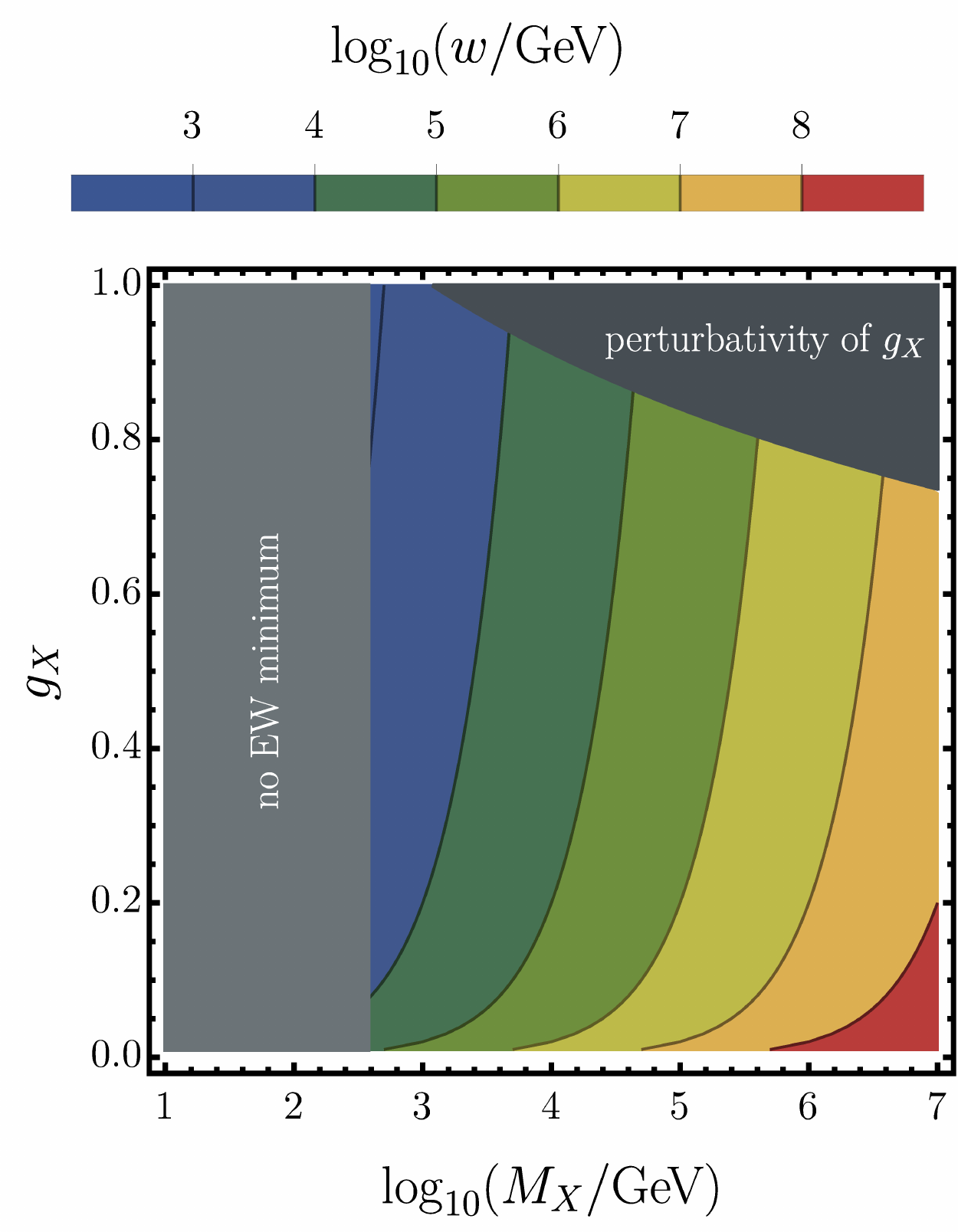}
\caption{Values of the new scalar mass $M_S$ (left panel) and the VEV $w$ (evaluated at $\mu=M_X$) (right panel). In the left panel the thick black line indicates where $M_S=M_H=125\g$ and across this line mass ordering between $S$ and $H$ changes (to the left of the line $M_S<M_H$, and to the right $M_H<M_S$). To the right of the dotted line $\xi_H$ becomes numerically equal to 1. The dashed lines indicate a discrepancy between the running and the pole mass (in percent). Grey-shaded regions are excluded.%, see caption of figure~\ref{fig:scalar-couplings}.
\label{fig:MS-w}}
\end{figure}

We present the result of the scan for $M_S$ and $w$ (the VEV of $\f$) in figure~\ref{fig:MS-w}. The new scalar $S$ is heavier than the Higgs boson in most of the parameter space. The dashed lines in the plot represent the disparity between the mass obtained by solving eq.~\eqref{eq:gap-scalar} iteratively and the mass estimated from the effective potential approximation. Although the differences are not negligible, they do not exceed 10\% even in the upper right region of the parameter space. Finally, the region of low $X$ masses is excluded because it is not possible the reproduce a stable minimum with the correct Higgs VEV and mass in this regime, while the upper right corner is cut off by the condition $g_X (M_Z) \leq 1.15$ for the perturbativity of the dark gauge coupling.

%%%%%%%%%%%%%%%%%%%%%%%%%%%%%%%%%%%%%%%%%%%%%%%
%%%%%%%%%%%%%%%%%%%%%%%%%%%%%%%%%%%%%%%%%%%%%%%
\section{Dark matter}
%%%%%%%%%%%%%%%%%%%%%%%%%%%%%%%%%%%%%%%%%%%%%%%
%%%%%%%%%%%%%%%%%%%%%%%%%%%%%%%%%%%%%%%%%%%%%%%

Our DM candidates are the three vector bosons $X_\mu^a$ (where $a = 1,\,2,\,3$) of the hidden sector gauge group $SU(2)$ with mass $M_X = \frac{1}{2} g_X w$. As discussed in~\cite{Gross:2015}, the gauge bosons are stable due to an intrinsic $\mathbb{Z}_2 \cross \mathbb{Z}_2'$ symmetry associated with complex conjugation of the group elements and discrete gauge transformations. This discrete symmetry actually generalizes to a custodial $SO(3)$~\cite{Hambye:2008} and the dark gauge bosons are degenerate in mass.

For the standard freeze-out mechanism, the Boltzmann equation has the form
%%%%%%%%%%%%%%%%%%%%%%%
\be
\frac{\dd n}{\dd t} + 3\,H\,n = - \, \frac{\left< \sigma v \right>_{\rm ann}}{3}  \left( n^2 - n^2_{eq} \right) - \frac{2\left< \sigma v \right>_{\rm semi}}{3} \, n \left( n - n_{eq} \right)\,.
\ee
The annihilation cross section is dominated by the $X X \rightarrow S S$ process%. The leading order term is 
%%%%%%%%%%%%%%%%%%%%%%%%%%%
\be 
\left< \sigma v \right>_{\rm ann} = \frac{11 g_X^4}{2304 \pi M^2_X}\,, %\frac{\cos^4\alpha}{M^2_X} \,.
\ee
%%%%%%%%%%%%%%%%%%%%%%%%%%%
while the semiannihilation cross section is dominated by the $X X \rightarrow X S$ process%. The leading order term is
%%%%%%%%%%%%%%%%%%%%%%%%%%%
\be 
\left< \sigma v \right>_{\rm semi} = \frac{3 g_X^4}{128 \pi M_X^2}\,. %\frac{\cos^2\alpha}{M_X^2} \,.
\ee
%%%%%%%%%%%%%%%%%%%%%%%%%%%
Interestingly, the semiannihilation processes dominate since $\left< \sigma v \right>_{\rm semi} \sim 5 \left< \sigma v \right>_{\rm ann}$.

Solving the Boltzmann equation, we obtain the dark matter relic abundance
%%%%%%%%%%%%%%%%%%%%%%%%%%%%
\be 
% \Omega_{X} h^2 = 3\times 
\Omega_{X} h^2 = \frac{1.04\times 10^9 \, \GeV^{-1}}{\sqrt{g_*} \, M_P \, J(x_f)}, \qquad J(x_f) = \int_{x_f}^\infty dx \, \frac{\left< \sigma v \right>_{\rm ann} + 2 \left< \sigma v \right>_{\rm semi}}{x^2}\,,
\label{eq:RelicDensity}
\ee
%%%%%%%%%%%%%%%%%%%%%%%%%%%%
%where the value of the freeze-out point can be obtained iteratively~\cite{Kolb:1990vq}:
%%%%%%%%%%%%%%%%%%%%%%%
%\be
%x_f = \ln \left\{ 0.038 \, \frac{3 M_X M_{P}}{\sqrt{g_*(T_f) x_f}} \Big[ \, c \, (c + 2) \left< \sigma v \right>_{\rm ann} + 2 \, c \, ( c + 1 ) %\left< \sigma v \right>_{\rm semi} \, \Big] \right\} . \label{eq:fzpoint}
%\ee
%%%%%%%%%%%%%%%%%%%%%%%
% the constant $c$ we use $c=1/2$~\cite{Griest:1990kh} and we find typical values between $x_f \approx 25-26$ and for the DM mass range that we consider in our numerical analysis. Finally, for $M_X \gg M_S$ and $\theta \ll 1$, the correct relic abundance ($\Omega_{\rm DM} h^2 = 0.120 \pm 0.001$~\cite{Planck:2018vyg}) is reproduced if 
where $x_f \approx 25-26$ and $x = M_X / T$. 
The correct relic abundance $\Omega_{\rm DM} h^2 = 0.120 \pm 0.001$ is reproduced if 
%%%%%%%%%%%%%%%%%%%%%%%%%%%%%%%%%%%%%
\be 
g_X \approx 0.9 \times \sqrt{\frac{M_X}{1 \ \rm TeV}}\,. %\quad \rm and \quad v_\phi \approx 2.2 \ \rm TeV \times \sqrt{\frac{M_X}{1 \ \rm TeV}}\,. 
\ee
%%%%%%%%%%%%%%%%%%%%%%%%%%%%%%%%%%%%%
Finally, DM particles can scatter off of nucleons, with the spin-independent cross section given by
%\small{
\be
\sigma_{\rm SI} = \frac{m_N^4  f^2 }{16\pi v^2}
 \bigg(\frac{1}{M_S^2} - \frac{1}{M_H^2}\bigg)^2 g_X^2 \sin^2 2\alpha
% \simeq \frac{f^2 g_X^2 m_N^4 }{4 \pi M_s^4 w^2} 
 \simeq  \frac{64\pi^3 f^2  m_N^4}{81 M_X^6} 
 \approx 0.6 \times 10^{-45}\ \rm{cm}^2  \left(\frac{\TeV}{M_X}\right)^6 \,.
\ee
%}
Then, to evade the experimental bounds we would have $\sigma_{\rm SI} < 1.5 \times 10^{-45}\ \rm{cm}^2 \, \left(M_X/\TeV\right)$ for $M_X > 0.88\TeV $.

%%%%%%%%%%%%%%%%%%%%%%%%%%%%%%%%%%%%%%%%%%%%%%%
%%%%%%%%%%%%%%%%%%%%%%%%%%%%%%%%%%%%%%%%%%%%%%%
\section{Finite temperature}
%%%%%%%%%%%%%%%%%%%%%%%%%%%%%%%%%%%%%%%%%%%%%%%
%%%%%%%%%%%%%%%%%%%%%%%%%%%%%%%%%%%%%%%%%%%%%%%
The temperature-dependent effective potential is
\be
V(h,\f,T)=\Vtree(h,\f)+\Vone(h,\f)+\VT(h,\f,T)+V_{\textrm{daisy}}(h,\f,T).\label{eq:V-eff}
\ee
The finite-temperature correction is
\be
\VT(h,\f,T)=\frac{T^4}{2\pi^2}\sum_{a} n_a J_a\left(\frac{M_a(h,\f)^2}{T^2}\right),\label{eq:thermal-pot}
\ee
where the sum runs over particle species. $J_a$ denotes the thermal function, which is given by
\be
J_{F,B}(y^2) =  \int_0^\infty \dd{x} x^2 \log(1\pm e^{-\sqrt{x^2 + y^2}}),\label{eq:thermal-functions}
\ee
where ``$+$'' for fermions ($J_F$) and ``$-$'' for bosons ($J_B$).
The correction from the daisy-resummed diagrams is
\be
V_{\textrm{daisy}}(h,\f,T)=-\frac{T}{12\pi}\sum_i n_i \left[(M_{i,\textrm{th}}^2(h,\f,T))^{3/2}-(M_i^2(h,\f))^{3/2}\right],
\ee
where $n_i$ is the number of degrees of freedom, $M_{i,\textrm{th}}$ denotes thermally corrected mass, and $M_i$ the usual field dependent mass.
%
% \begin{figure}[h!t]
%     \centering
%     \includegraphics[height=.2\textheight]{VatTnuc-barrier}\hspace{6pt}
%     \includegraphics[height=.2\textheight]{VatTnuc-min}
%     \caption{RG-improved effective potential at nucleation temperature along the $\f$ direction for a benchmark point with $g_X=0.9$, $M_X=10^4\g$ in two ranges of the field values -- around the barrier, where the tunnelling takes place (left panel), and around the minimum (right panel).}
%     \label{fig:VatTnuc}
% \end{figure}

% \be
% \mu=\textrm{max}\left(\frac{1}{2}g_X(M_X)\f,\  0.1\g\right)\equiv \textrm{max}\left(\overline{M}_X(\f),\ 0.1\g\right).\label{eq:choice-of-mu}
% \ee
The zero-temperature part of the effective potential along the $\varphi$ direction reads
\be
V(\f)=\frac{1}{4}\lc(t) Z_{\f}(t)^2\f^4+\frac{9M_X(\f,t)^4}{64\pi^2}\left(\log\frac{M_X(\f,t)^2}{\mu^2}-\frac{5}{6}\right),\label{eq:V-one-loop-along-phi}
\ee
where $t=\log\frac{\mu}{\mu_0}$, $\mu_0=M_Z$, $M_X(\f,t)=\frac{1}{2}g_X(t)\sqrt{Z_{\f}(t)}\f$, $\mu = \frac{1}{2} g_X (M_X) \varphi \equiv \overline{M}_X(\f)$.

%Note that our approach for the renormalization-group improvement of the effective potential differs from the ones usually employed in the literature. 
Note that we include more terms in the renormalisation-group improved potential than in the approaches often found in the literature.
In detail:

\begin{enumerate}
    \item The approach of~\cite{Hambye:2018, Baldes:2018} approximates the running quartic coupling via its $\beta$ function, relates the renormalisation scale with the field and uses as a reference scale the scale at which $\lambda_\varphi$ changes sign,
    \be
    V_1\approx \frac{1}{4}\lc(t) \f^4 \approx \frac{1}{4}\frac{9 g_X^4}{128\pi^2}\log(\frac{\f}{\f_0}),\label{eq:potential-Strumia}
    \ee
    where $t=\log\frac{\mu}{\f_0}$, $\lambda_\varphi(0)=0$ and $g_X$ is evaluated at $\mu=\f_0$ (the running of $g_X$ is not included).

    \item The approach of~\cite{Ellis:2020} also approximates the one-loop potential by the tree-level potential with running coupling but uses $\mu=\f$ and some fixed reference scale $\mu_0 = m_t$,
    \be
    V_2\approx \frac{1}{4}\lc(t) \f^4,\label{eq:potential-updated}
    \ee
    where $t=\log(\frac{\f}{\mu_0})$.
\end{enumerate}

To better understand which contributions are crucial we perform a series of approximations or modifications on our approach, the results of which are presented in the right panel of figure~\ref{fig:potential-comparison}. Namely:

\begin{figure}[ht]
    \centering
    \includegraphics[height=.23\textheight]{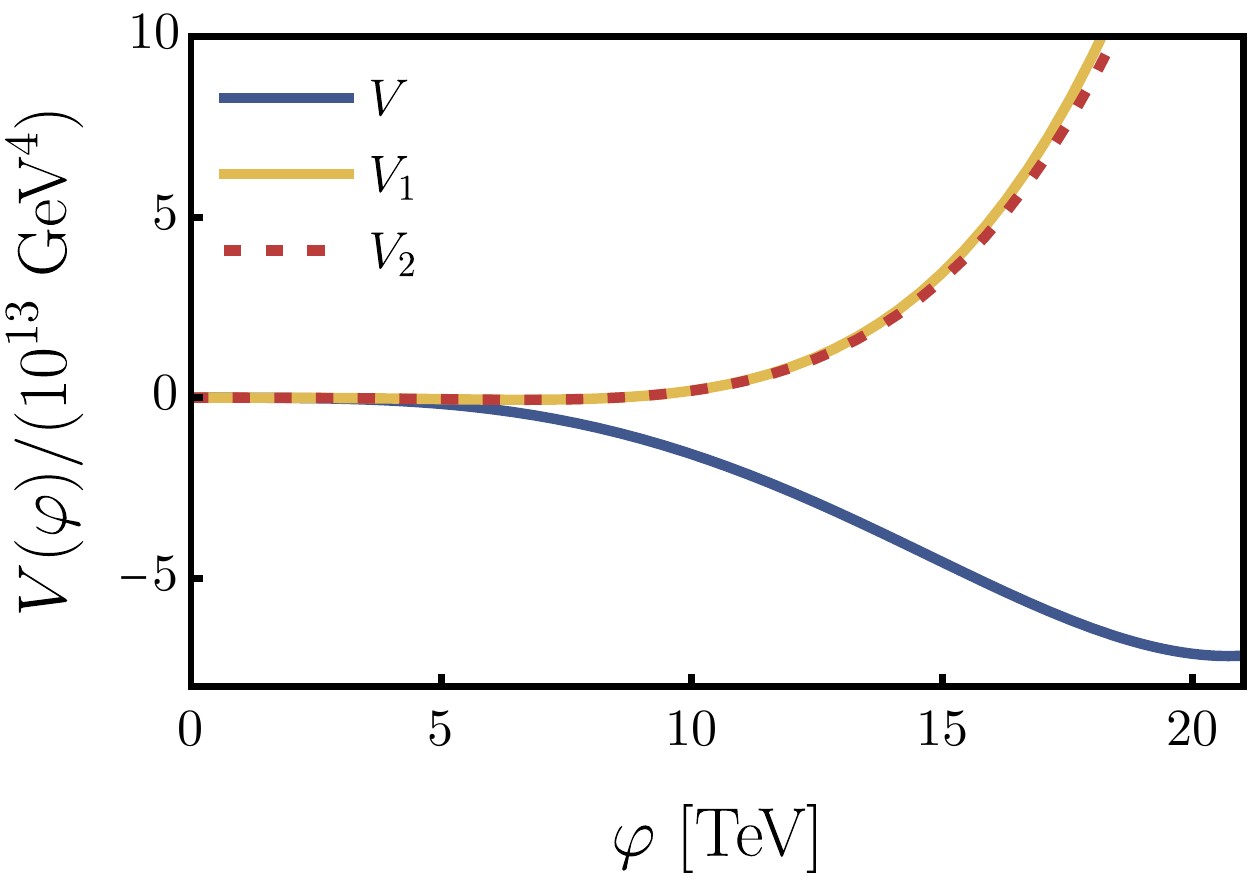}\hspace{2pt}
    \includegraphics[height=.23\textheight]{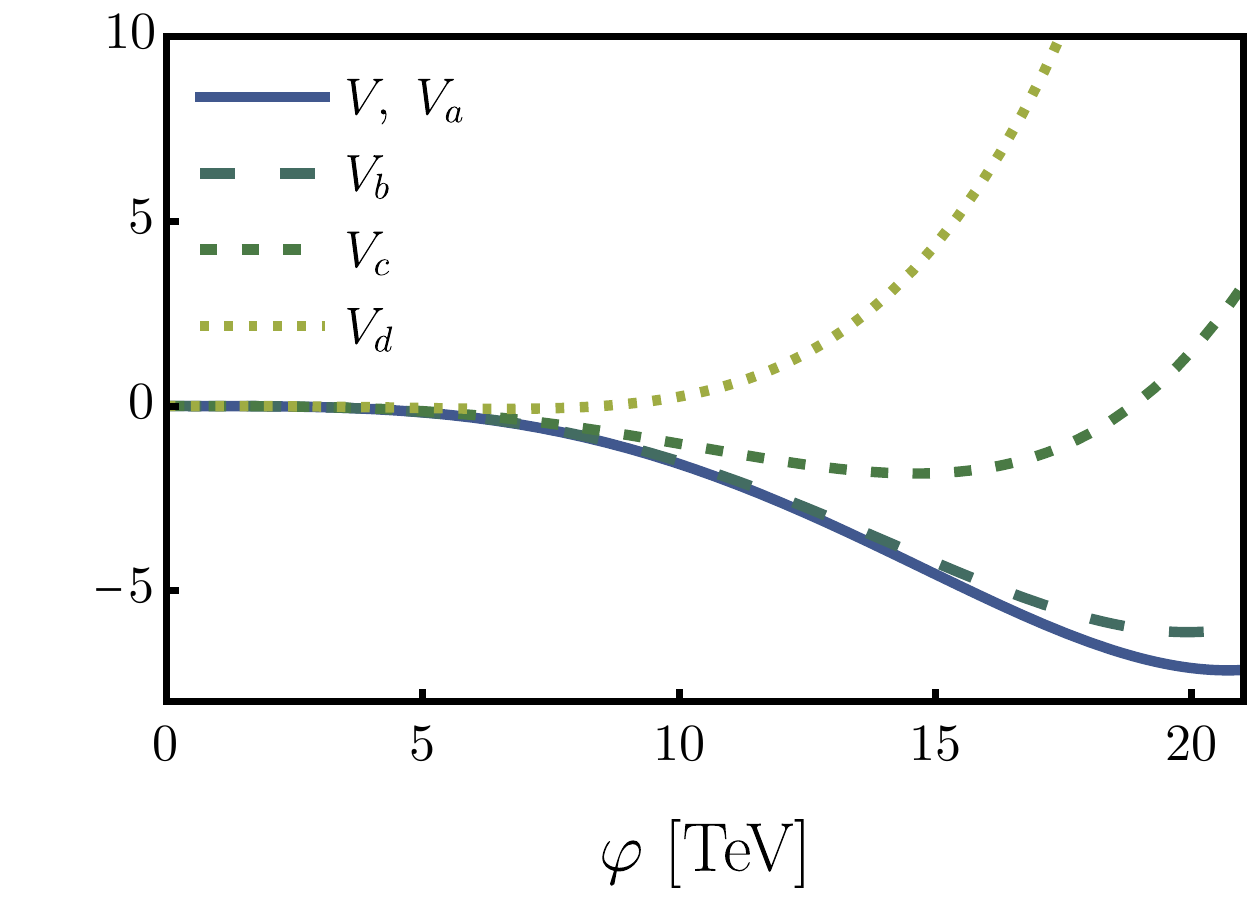}
    \caption{Effective potential at zero temperature along the $\f$ direction for a benchmark point with $g_X=0.9$, $M_X=10^4\g$ (defined at $\mu=M_X$). Left panel: Comparison of different approaches used in the literature, $V_1$ of eq.~\eqref{eq:potential-Strumia} (yellow solid), $V_2$ of eq.~\eqref{eq:potential-updated} (dashed red) and the full potential $V$ of eq.~\eqref{eq:V-one-loop-along-phi} used in this work (solid blue). Right panel: Comparison of different approximations imposed on the full potential $V$ of eq.~\eqref{eq:V-one-loop-along-phi} used in this work (solid blue) discussed in the main text: $V_b$ (long-dashed darkest green), $V_c$ (medium-dashed dark green), $V_d$ (short-dashed light green).}
    \label{fig:potential-comparison}
\end{figure}

\begin{enumerate}%[label=(\alph*)]
    \item $V_a$ corresponds to the potential $V$ with the part proportional to the logarithm neglected. $V_a$ exactly overlaps with the full potential (solid blue line).
    \item $V_b$ corresponds to the potential $V$ with the choice of $\mu=\f$ (darkest green, long-dashed line). This choice alone does not modify the potential significantly with respect to our choice (solid blue line).
    \item $V_c$ corresponds to the potential $V$ with the constant $-\frac{5}{6}$ neglected (dark green, medium-dashed curve). Since the omission of the logarithm (with our choice of the scale) does not visibly modify the result, $V_c$  is equivalent to using the tree-level part of $V$. Here the difference with respect to the full potential is significant. It is understandable, since the choice of the scale was such as to get rid of the logarithmic term but not the $\frac{5}{6}$ constant.
    \item $V_d$ corresponds to the tree-level part of $V$ but with the choice $\mu=\f$ (light green, short-dashed line), which makes this choice very close to $V_1$ and $V_2$ discussed above. Clearly, $V_d$ differs significantly from the full potential.
\end{enumerate}

% \be
% g\f\sim T.\label{eq:finite-T-scaling}
% \ee

%%%%%%%%%%%%%%%%%%%%%%%%%%%%%%%%%%%%%%%%%%%%%%%
%%%%%%%%%%%%%%%%%%%%%%%%%%%%%%%%%%%%%%%%%%%%%%%
\section{Phase transition and gravitational wave signal}
%%%%%%%%%%%%%%%%%%%%%%%%%%%%%%%%%%%%%%%%%%%%%%%
%%%%%%%%%%%%%%%%%%%%%%%%%%%%%%%%%%%%%%%%%%%%%%%

A first-order phase transition proceeds through nucleation, growth and percolation of bubbles filled with the broken-symmetry phase in the sea of the symmetric phase. This corresponds to the fields tunnelling through a potential barrier. In our case, we have checked that tunnelling proceeds along the $\f$ direction, while the transition in the $h$ direction is smooth. 

\subsection{Important temperatures}

The temperatures relevant to our discussion are:

\paragraph{Critical Temperature $T_c$.} At high temperatures the symmetry is restored and the effective potential has a single minimum at the origin of the field space. As the Universe cools down, a second minimum is formed. At the critical temperature, the two minima are degenerate, and for lower temperatures, the minimum with broken symmetry becomes the true vacuum. This is the temperature at which the tunnelling becomes possible.

\paragraph{Thermal Inflation Temperature $T_V$.}

\begin{figure}[h!t]
\center
\includegraphics[width=.8\textwidth]{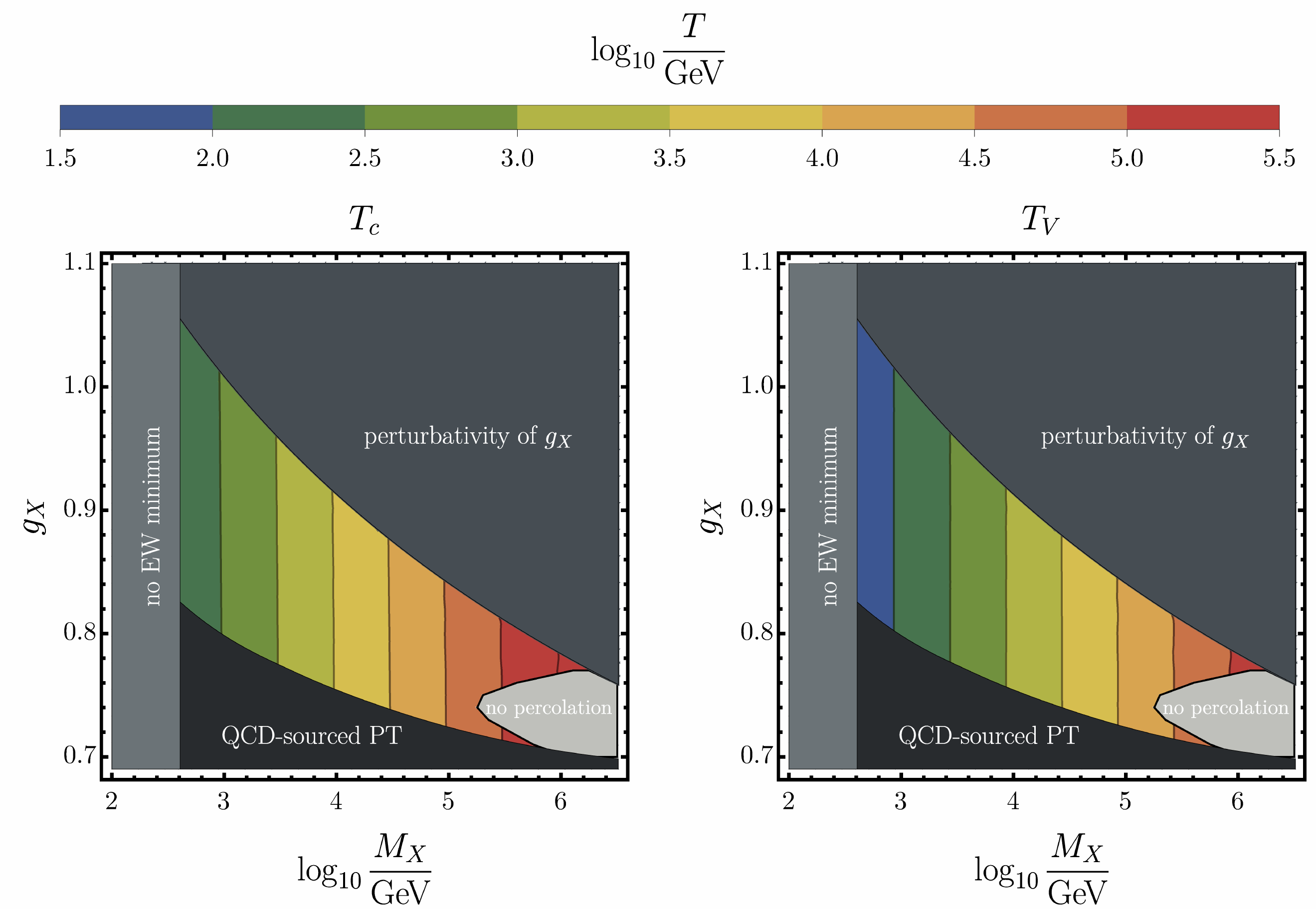}
\caption{The values of the critical temperature $T_c$ (left panel) and the temperature at which thermal inflation starts $T_V$ (right panel). \label{fig:Tc-TV}}
\end{figure}

If there is large supercooling, i.e.\ the phase transition is delayed to low temperatures, much below the critical temperature, it is possible that a period of thermal inflation due to the false vacuum energy appears before the phase transition completes. The Hubble parameter can be written as
\be
H^2=\frac{1}{3\bar{M}_{\rm{Pl}}^2}(\rho_R+\rho_V)=\frac{1}{3\bar{M}_{\rm{Pl}}^2}\left(\frac{T^4}{\xi^2_g}+\Delta V\right), \quad \xi_g=\sqrt{30/(\pi^2 g_*)}\,,
\ee
where $\Delta V$ is the difference between the values of the effective potential at false and true vacuum. The onset of the period of thermal inflation can be approximately attributed to the temperature at which vacuum and radiation contribute to the energy density equally, 
\be
 %\frac{T_V^4}{\xi^2_g} = \Delta V
 T_V \equiv \left(\xi^2_g \Delta V \right)^\frac{1}{4}.
\ee
For supercooled transitions, it is a good approximation to assume that $\Delta V$ is independent of the temperature below $T_V$.
By using the temperature $T_V$, the Hubble constant can be rewritten as
\be
H^2 \simeq \frac{1}{3\bar{M}_{\rm{Pl}}^2 \xi^2_g}\left(T^4+T_V^4\right).
\ee
In the case of large supercooling, the contribution to the Hubble parameter from radiation energy can be neglected, leaving
\be
H^2 \simeq H^2_V=\frac{1}{3\bar{M}_{\rm{Pl}}^2}\Delta V.
\ee
In figure~\ref{fig:Tc-TV} there are the same excluded areas as before and two new shaded regions. The lower left corner (darkest grey) is not analysed because there the PT is sourced by the QCD phase transition, which is beyond the scope of the present work. The light-grey region around $M_X\approx 10^6\g$ is where the percolation criterion of eq.~\eqref{eq:percolation-crit} is violated and is discussed in more detail below.
%

% \be
% H^2=\frac{1}{3\bar{M}_{\rm{Pl}}^2}(\rho_R+\rho_V)=\frac{1}{3\bar{M}_{\rm{Pl}}^2}\left(\frac{T^4}{\xi^2_g}+\Delta V\right),
% \ee

% \be \label{eq:T-thermal-inflation}
%  %\frac{T_V^4}{\xi^2_g} = \Delta V
%  T_V \equiv \qty(\xi^2_g \Delta V )^\frac{1}{4}.
% \ee

% \be
% H^2 \simeq \frac{1}{3\bar{M}_{\rm{Pl}}^2 \xi^2_g}\left(T^4+T_V^4\right).
% \ee

% \be
% H^2 \simeq H^2_V=\frac{1}{3\bar{M}_{\rm{Pl}}^2}\Delta V.
% \ee

\paragraph{Nucleation Temperature $T_n$.}

Below the critical temperature, nucleation of bubbles of true vacuum becomes possible. To compute the decay rate of the false vacuum we start by solving the bounce equation,
\be
\frac{\dd^2 \f}{\dd r^2}+\frac{2}{r}\frac{\dd \f}{\dd r}=\frac{\dd V(\f,T)}{\dd \f}, \qquad \frac{\dd \f}{\dd r}=0 \quad \text{ for } \quad r=0 \quad \text{ and } \quad \f\to0 \quad \text{ for } \quad r\to\infty.
\ee
Once the bubble profile is known we can compute the Euclidean action along the tunnelling path
\be
S_3(T)=4\pi\int r^2\mathrm{d}r \frac{1}{2}\left(\frac{\mathrm{d}\f}{\mathrm{d}r}\right)^2 + V(\f,T).
\ee
Then the decay rate of the false vacuum due to the thermal fluctuations is given by
\be
\Gamma(T)\approx T^4\left(\frac{S_3(T)}{2\pi T}\right)^{3/2} e^{-S_3(T)/T}.
\ee
The nucleation temperature is defined as the temperature at which at least one bubble is nucleated per Hubble volume, which can be interpreted as the onset of the PT.  
\be
N(T_n)=1=\int_{t_c}^{t_n}dt\frac{\Gamma(t)}{H(t)^3}=\int_{T_n}^{T_c}\frac{dT}{T}\frac{\Gamma(T)}{H(T)^4}.
\ee
The common criterion for evaluating $T_n$ as $S_3/T_n\approx 140$ is not reliable in the case of strongly supercooled transitions.

% \be
% \frac{\mathrm{d}^2\f}{\mathrm{d}r^2}+\frac{2}{r}\frac{\mathrm{d}\f}{\mathrm{d}r}=\frac{\mathrm{d}V(\f,T)}{\mathrm{d}\f},
% \ee

% \be
% S_3(T)=4\pi\int \mathrm{d}r ~r^2 \left[ \frac{1}{2}\left(\frac{\mathrm{d}\f}{\mathrm{d}r}\right)^2 + V(\f,T) \right].
% \ee

% \be
% \Gamma(T)\approx T^4\left(\frac{S_3(T)}{2\pi T}\right)^{3/2} e^{-S_3(T)/T}.
% \ee

% \be
% N(T_n)=1=\int_{t_c}^{t_n}dt\frac{\Gamma(t)}{H(t)^3}=\int_{T_n}^{T_c}\frac{dT}{T}\frac{\Gamma(T)}{H(T)^4}.
% \ee

\paragraph{Percolation Temperature $T_p$.}

When the bubbles of the true vacuum percolate, most of the bubble collisions take place. Therefore, the percolation temperature is the relevant temperature for the GW signal generation. The probability of finding a point still in the false vacuum at a certain temperature is given by $P(T) = e^{-I(T)}$, where $I(T)$ is the amount of true vacuum volume per unit comoving volume and reads as 
\be
		I(T) = \frac{4\pi}{3} \int_{T}^{T_c} \dd{T^\prime} \frac{\Gamma(T^\prime)}{T^{\prime 4} H(T')}\left(\int_{T}^{T'} \frac{\dd{\tilde T} }{ H(\tilde T)}\right)^3.
\ee
We can distinguish between the vacuum and radiation domination period which leads to the Hubble parameter in the following form:
\begin{align} 
	H(T) \simeq 
\left\{\begin{array}{ll}
	H_{\mathrm{R}}(T)=\frac{T^{2}}{\sqrt{3} \bar{M}_{\mathrm{pl}} \xi_{g}}, &\quad \mathrm{for}\quad T > T_{V}, \\
	H_{\mathrm{V}}=\frac{T_{V}^{2}}{\sqrt{3} \bar{M}_{\mathrm{pl}} \xi_{g}}, &\quad \mathrm{for}\quad T < T_{V}.
\end{array}\right.
\end{align}
We can thus write a simplified version of $I(T)$ valid in the region where $T<T_V$:
\begin{align}
\begin{split}
I_{\mathrm{RV}}(T) =\frac{4 \pi}{3 H_{\mathrm{V}}^4}\left(\int_{T_V}^{T_c} \frac{d T^{\prime} \Gamma\left(T^{\prime}\right)}{T^{\prime 6}} T_V^2\left(2 T_V-T-\frac{T_V^2}{T^{\prime}}\right)^3+\int_T^{T_V} \frac{d T^{\prime} \Gamma\left(T^{\prime}\right)}{T^{\prime}}\left(1-\frac{T}{T^{\prime}}\right)^3\right)
\end{split}
\end{align}

The percolation criterion is given by%~\cite{Vinod:1971}
\be
I_{\mathrm{RV}}(T_p)= 0.34\,, \qquad \text{or} \qquad P(T_p) = 0.7 \,.
\label{eq:percolation-temp}
\ee
The fraction 0.34 is the ratio of the volume in equal-size and randomly-distributed spheres (including overlapping regions) to the total volume of space for which percolation occurs in three-dimensional Euclidean space, and implies that at $T_p$ at least 34\% of the (comoving) volume is converted to the true minimum.

\begin{figure}[h!t]
\center
\includegraphics[width=.8\textwidth]{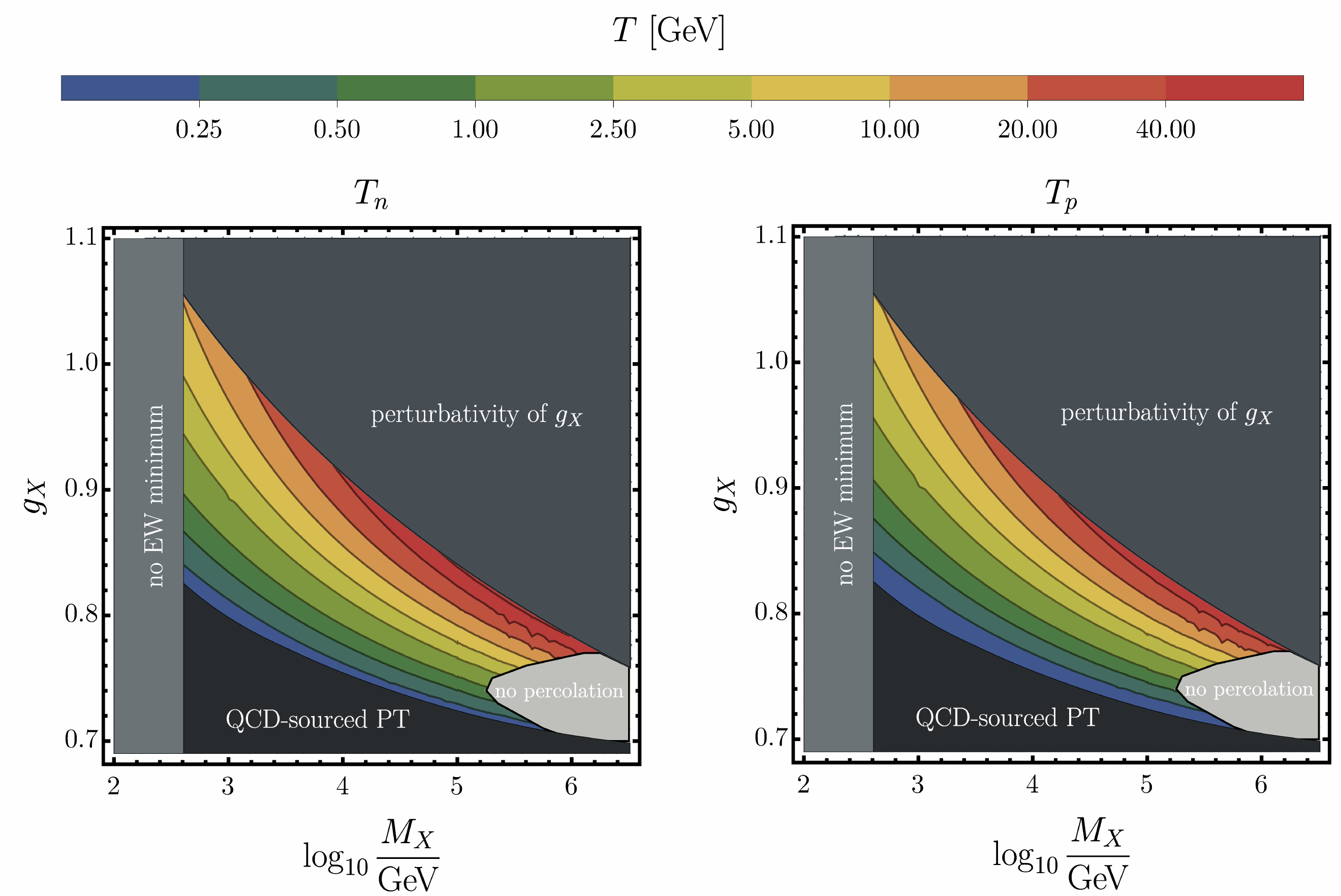}
\caption{The values of the nucleation temperature $T_n$ (left panel) and the percolation temperature $T_p$ (right panel).\label{fig:Tn-Tp}}
\end{figure}

Comparing to the values of $T_n$ (fig.~\ref{fig:Tn-Tp}) one can see that these two temperatures are of the same order, yet they differ, hence one should not use $T_n$ as a proxy for the temperature at which the PT proceeds in case of the models with large supercooling.

One also needs to make sure that the volume of the false vacuum $V_f \sim a^3(T) P(T)$ is  decreasing around the percolation temperature. This condition is especially constraining in models featuring strong supercooling, as thermal inflation can prevent bubbles from percolating. It can be expressed as
\begin{align}
\frac{1}{V_f} \frac{\dd V_f}{\dd t} = 3 H(t) - \frac{\dd I(t)}{\dd t} = H(T)\left(3 + T \frac{\dd I(T)}{\dd T} \right) < 0.
\label{eq:percolation-crit}
\end{align}

%

% \begin{align}
% 		I(T) = \frac{4\pi}{3} \int_{T}^{T_c} \dd{T^\prime} \frac{\Gamma(T^\prime)}{T^{\prime 4} H(T')}\qty(\int_{T}^{T'} \frac{\dd{\tilde T} }{ H(\tilde T)})^3.
% \end{align}

% \begin{align} \label{hubble_approx}
% 	H(T) \simeq 
% \left\{\begin{array}{ll}
% 	H_{\mathrm{R}}(T)=\frac{T^{2}}{\sqrt{3} \bar{M}_{\mathrm{pl}} \xi_{g}}, &\quad \mathrm{for}\quad T > T_{V}, \\
% 	H_{\mathrm{V}}=\frac{T_{V}^{2}}{\sqrt{3} \bar{M}_{\mathrm{pl}} \xi_{g}}, &\quad \mathrm{for}\quad T < T_{V}.
% \end{array}\right.
% \end{align}

% \begin{align}
% \begin{split}
% I_{\mathrm{RV}}(T) =\frac{4 \pi}{3 H_{\mathrm{V}}^4}\left(\int_{T_V}^{T_c} \frac{d T^{\prime} \Gamma\left(T^{\prime}\right)}{T^{\prime 6}} T_V^2\left(2 T_V-T-\frac{T_V^2}{T^{\prime}}\right)^3+\int_T^{T_V} \frac{d T^{\prime} \Gamma\left(T^{\prime}\right)}{T^{\prime}}\left(1-\frac{T}{T^{\prime}}\right)^3\right) .
% \end{split}
% \end{align}

% \be
% I_{\rm RV}(T_p) = 0.34.\label{eq:percolation-temp}
% \ee

% \begin{align}
% \frac{1}{V_f} \dv{V_f}{t} = 3 H(t) - \dv{I(t)}{t} = H(T)\qty(3+T\dv{I(T)}{T}) < 0.\label{eq:percolation-crit}
% \end{align}

\paragraph{Reheating Temperature $T_r$.}

%We assume that 
At the end of the phase transition, the Universe is in a vacuum-dominated state. Then the total energy released in the phase transition is $\Delta V(T_p)\approx \Delta V (T=0)\equiv \Delta V$. If reheating is instantaneous, this whole energy is turned into the energy of radiation,
\be
\Delta V = \rho_R(T_r)=\rho_R(T_V) \qquad \rightarrow \qquad T_r=T_V.
\ee

On the other hand, if at $T_p$ the rate of energy transfer from the $\f$ field to the plasma, $\Gamma_{\f}$, is smaller than the Hubble parameter, $\Gamma_{\f} < H(T_p)$, then the energy will be stored in the scalar field oscillating about the true vacuum and redshift as matter until $\Gamma_{\f}$ becomes comparable to the Hubble parameter. In this case 
\be
T_r=T_V\sqrt{\frac{\Gamma_{\f}}{H_*}}.
\ee

%In order to determine which of these scenarios takes place one has to evaluate $\Gamma_{\f}$. Before the phase transition the energy is stored in the $\f$ field. Thus, for assessing the efficiency of reheating we should know the decay rate of the $\f$ field which quantifies the energy transfer rate from $\f$ to the plasma. The $\f$ field can be understood as a mixture of the mass eigenstates $H$ and $S$ (by the inverse of eq.~\eqref{eq:mixing}) for which the decay widths are well defined. The decay width of $H$ is equal to the SM Higgs decay rate rescaled due to the mixing by $\xi_H^2$ (see eq.~\eqref{eq:xi}). Since the SM Higgs decay width is fairly large, we can safely assume that the $H$ component of $\f$ decays quickly. Due to the mixing between $h$ and $\f$ the decay width of $S$ is a sum of SM-like decays with couplings rescaled by $\xi_S$ and, in case $M_S>2M_H$ the scalar decay $S\to HH$. The decay to a pair of gauge bosons $S\to XX$ is kinematically forbidden. Therefore, 
The rate of energy transfer from $\f$ to the plasma reads
\begin{align}
    \Gamma_{\f}=& %(1-\xi_S^2)\Gamma(S)=
    \xi_S^2(1-\xi_S^2)\Gamma_{\mathrm{SM}}(S)+(1-\xi_S^2)\Gamma(S\to HH) ,\quad
\xi_S=\left\{\begin{array}{lrl}
    -\sin\theta & \textrm{ for }& M_H\leqslant M_S  \\
    \cos\theta & \textrm{ for }& M_H>M_S
\end{array}\right.
\end{align}
where $\Gamma_{\mathrm{SM}}$ denotes a decay width computed as in the SM, i.e. with the same couplings and decay channels, but for a particle of mass $M_S$. %From the above formula it is clear that if there is no mixing between the scalars the only available decay channel is the scalar one, $S\to HH$. It could be suspected that, since the mixing between the two scalars is small, the approximation of no mixing should hold. However, it turns out that 
The mixing enhances the decay width twofold, first, it amplifies the coupling $SHH$ as compared to $\f h h$ and, moreover, it allows a contribution from the SM sector, which is especially important when the $S\to HH$ decay is kinematically forbidden.

% \be
% \Delta V = \rho_R(T_r)=\rho_R(T_V),\label{eq:condition-Tr-1}
% \ee

% \be
% T_r=T_V.\label{eq:Treh=TV}
% \ee

% \be
% T_r=T_V\sqrt{\frac{\Gamma_{\f}}{H_*}}  .
% \label{eq:Treh-general}
% \ee

% \begin{align}
%     \Gamma_{\f}=& (1-\xi_S^2)\Gamma(S)=\xi_S^2(1-\xi_S^2)\Gamma_{\mathrm{SM}}(S)+(1-\xi_S^2)\Gamma(S\to HH),\label{eq:gamma-phi}
% \end{align}

\begin{figure}
    \centering
    \includegraphics[width=.5\textwidth]{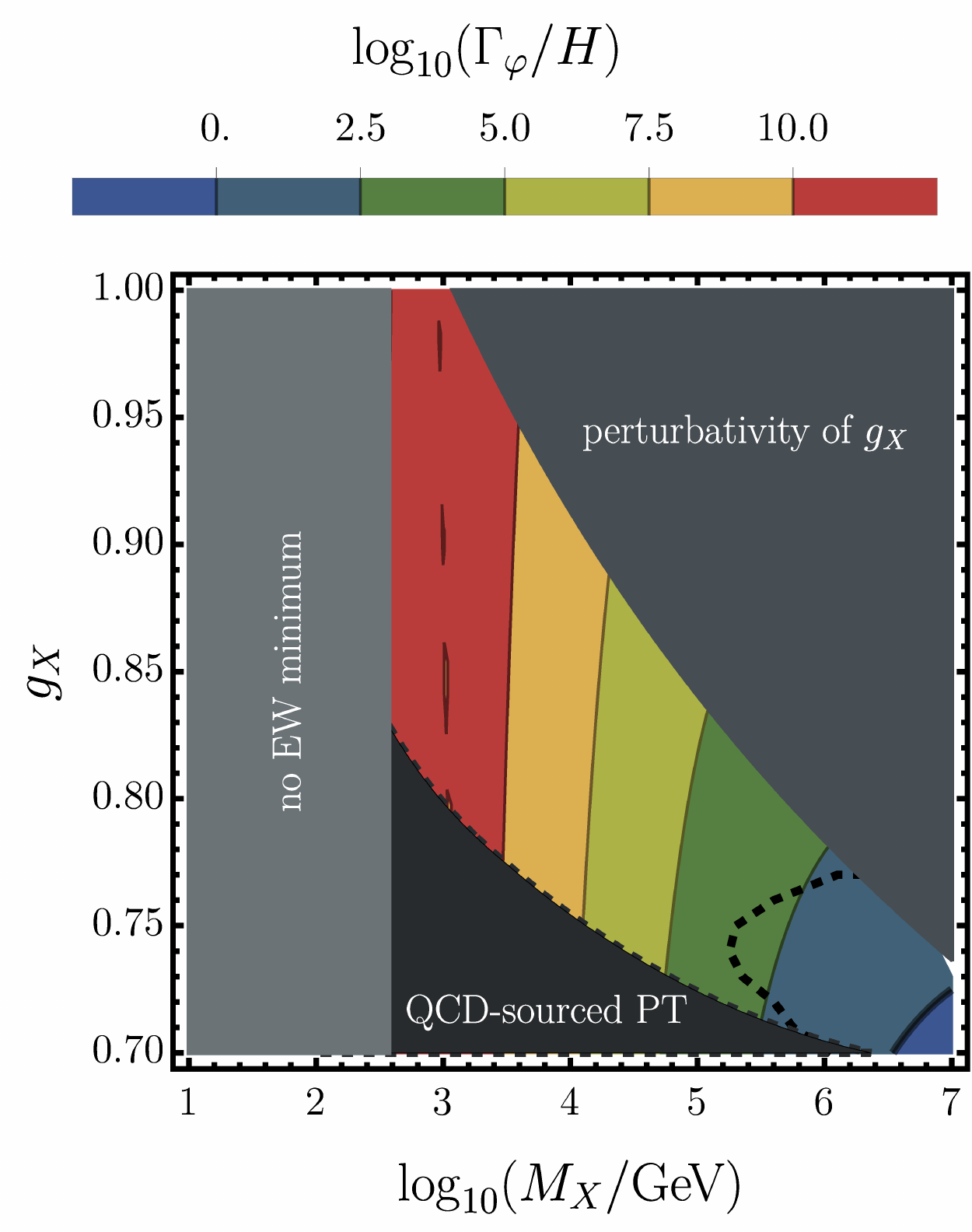}%\hspace{2pt}
    \caption{Contour plot of the decimal logarithm of the ratio of the energy transfer rate $\Gamma_{\f}$ to the Hubble parameter $H$. The equality $H=\Gamma_{\f}$ is indicated as a thick black solid line in the lower right corner. The percolation bound is shown as a black dashed line (in other plots it is shown as a light-grey region).
    \label{fig:reheating}}
\end{figure}

\subsection{Supercool Dark Matter?}

The authors of~\cite{Hambye:2018, Baldes:2018, Marfatia:2020} claim that for a wide range of parameters, there can be supercool DM. Their main assumptions are:
\begin{itemize}
    \item The true vacuum has zero energy, the energy in the false vacuum is $\Delta V\simeq 9m^4_\chi/(128 \pi^2)$, which implies that supercooling starts at
    \[
    T_{V}\simeq \frac{M_X}{8.5}\quad\text{and}\quad
    H_* = \sqrt{\frac{3}{\pi}}\,\frac{M^2_X}{4M_{\rm pl}}\,.
    \]
    \item Nucleation occurs when $S_3(T_n)/T_n\simeq 4 \ln (M_{\rm pl}/m_\chi)\simeq 142$.
    \item The reheating temperature is related to the thermal inflation temperature as \\
    $
    T_{\rm r} = T_{V} \ \min\left(1, \Gamma/H \right)^{1/2}$\,, where $\Gamma \simeq \Gamma_h \sin^2(v/w)$\,, with
    $\Gamma_h \approx 4 \ \rm{MeV}\,.
    $
    \item The DM abundance resulting from inflationary supercooling is 
    \[
    Y_{\rm DM} \equiv \frac{n_{\rm DM}|_{T=T_{\rm r}}}{s|_{T=T_{\rm r}}}= \frac{45 g_{\rm DM}}{2\pi^4 g_*}\frac{T_{\rm r}}{T_{\rm V}} \left( \frac{T_{\rm n}}{T_{\rm V}} \right)^3\,.
    %~~~~~~
    %Y^{\rm eq}_{\rm DM}= \frac{45 g_{\rm DM}}{2\pi^4 g_*}\,,
    \]
    \item For $T_{\rm r} < T_{\rm dec} \simeq M_X / 25 $, both supercooling and sub-thermal production contribute to the DM relic abundance,
    \[
    \Omega_{\rm DM}h^2=\Omega_{\rm DM}h^2|_{\rm supercool}
    +\Omega_{\rm DM}h^2|_{\rm sub-thermal}\,.
    \]
    \item For $T_{\rm r} > T_{\rm dec}$, the plasma thermalizes again,  and the usual freeze-out mechanism yields the relic abundance,
    \[
    \Omega_{\rm DM}h^2=\Omega_{\rm DM}h^2|_{\rm freeze-out}\,.
    \]

%The supercool DM population is what remains after the late time inflation. The gauge bosons are originally massless and their abundance gets suppressed by the dilution due to thermal inflation. In addition, after reheating, there is a sub-thermal population which can be produced through the thermal bath via scattering effects.

\end{itemize}

%%%%%%%%%%%%%%%%%%%%%%%%%%%%%%%%%%%%%
\begin{figure}
    %\centering
    \center
\includegraphics[width=.9\textwidth]{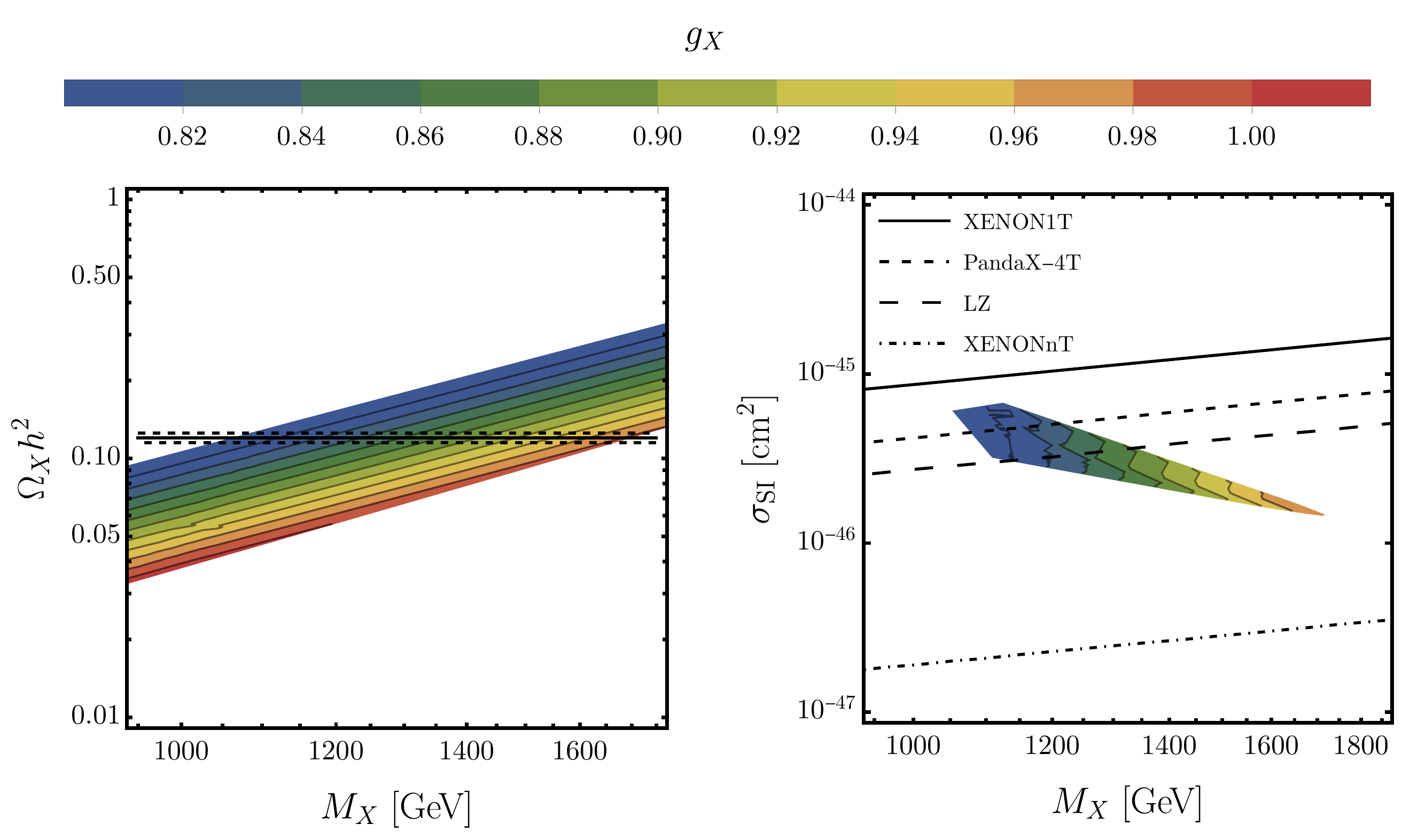}
    \caption{Left: Dark matter relic abundance $\Omega_X h^2$ with colour changing according to the value of the gauge coupling $g_X$. The black lines correspond to the measured value $\Omega_{\rm DM} h^2 = 0.120 \pm 5 \sigma$. Right: The spin-independent dark matter-nucleon cross section. The coloured region corresponds to points that reproduce the measured relic abundance within $5 \sigma$. The lines represent the exclusion limit from the XENON1T 2018~\cite{XENON:2018voc} (solid), PandaX-4T 2021~\cite{PandaX-4T:2021bab} (dashed), LZ 2022~\cite{LZ:2022ufs} (large dashed) and the scheduled XENONnT~\cite{XENON:2020kmp} (dot dashed) experiments.}
    \label{fig:relicplot}
\end{figure}
%%%%%%%%%%%%%%%%%%%%%%%%%%%%%%%%%%%%

Nevertheless, our analysis suggests that due to the percolation criterion which excludes $M_X$ above $\sim 10^6 \ \rm \GeV$ and the fact that $\Gamma_\varphi > H(T_p)$ in the rest of the DM range, we find $T_r > T_{\rm dec}$ for all parameter points. Hence, the supercool DM population gets diluted away, the sub-thermal population reaches thermal equilibrium again, and the relic abundance is produced as in the standard freezeout scenario (see fig.~\ref{fig:relicplot}). Our conclusions were also validated in a recent paper~\cite{Frandsen:2022klh}.

% New Paper (2301.00041): 
% \begin{itemize}
% \item ``Recently, the $SU(2)_X$ model was discussed in [2210.07075], and we find that their results for the $\alpha$ and $T_n$ parameters agree with our findings."
% \item ``We find that throughout the parameter space of interest in this work ... the DM abundance is obtained via the usual freeze-out mechanism."
% \end{itemize}

%Stochastic gravitational wave background produced by the first-order phase transition can be associated with three sources: bubble collisions, sound waves and turbulence in the plasma. The signal produced by the turbulence remains a subject of on-going discussion in the community, and its contribution is still burdened with large uncertainties. Therefore, we neglect it in this work.

\subsection{Gravitational waves}

The GW signal in the model under consideration can be sourced by bubble collisions. The spectrum is:
\be  
\Omega_{\text{col}}(f) = \left(\frac{R_*H_*}{5}\right)^2 \left( \frac{\kappa_{\text{col}} \alpha }{1+\alpha} \right)^2 S_{\textrm{col}}(f)\,. 
\ee
where $R_*$ is the length scale of the transition, $\kappa_{\text{col}}$ is the energy transfer efficiency factor at the end of the transition and $\alpha = \Delta V / 
 \rho_R (T_p)$ is the transition strength. The spectral shape $S_{\textrm{col}}$ and peak frequency are defined as
\be
S_{\textrm{col}} = 25.09  \left[ 2.41 \left(\frac{f}{ f_{\textrm{col}} }\right)^{-0.56} + 2.34 \left(\frac{f}{ f_{\textrm{col}} }\right)^{0.57} \right]^{-4.2}\,, \qquad f_{\textrm{col}} \simeq 0.13 \left(\frac{5}{R_*H_*}\right).
\ee
The spectra of the sound-wave-sourced GW are expressed as:
\be
\Omega_{\textrm{sw}}(f) = \left(\frac{R_*H_*}{5}\right) \left( 1 - \frac{1}{\sqrt{1+2\tau_{\textrm{sw}}H_* }} \right) 
\left( \frac{\kappa_{\text{sw}} \alpha }{1+\alpha} \right)^2 S_{{\textrm{sw}}}(f),
\ee
with
\be
S_{\textrm{sw}}(f) =\left(\frac{f}{f_{\textrm{sw}}}\right)^3 \left[\frac{4}{7}+\frac{3}{7}\left(\frac{f}{f_{\textrm{sw}}}\right)^2\right]^{7/2}\,,
\ee
where the duration of the sound wave period normalised to Hubble and the peak frequency can be expressed as
\be
\tau_{\textrm{sw}} H_* = \frac{R*H_*}{U_f}, \quad U_f \simeq \sqrt{\frac{3}{4}\frac{\alpha}{1+\alpha}\kappa_{\textrm{sw}}}\,, \quad f_{\textrm{sw}} \simeq 0.54 \left(\frac{5}{R_*H_*}\right)\,.
\ee

To assess the observability of a signal we compute the signal-to-noise (SNR) ratio for the detectors that have the best potential of observing the predicted signal, i.e.\ LISA and AEDGE. We calculate the SNR using the usual formula \cite{Caprini:2019egz, Robson:2018ifk}:

\begin{align} 
\mbox{SNR} = \sqrt{
\mathcal{T} 
\int_{f_{\rm{min}}}^{f_{\rm{max}}}
\dd f \left[ \frac{h^2\Omega_{\rm{GW}}(f)}{h^2\Omega_{\rm{Sens}}(f)} \right]^2
},
\end{align}
where $\mathcal{T}$ is the duration of collecting data and $h^2\Omega_{\rm{Sens}}(f)$ is the sensitivity curve of a given detector.
For calculations we have used data collecting durations as $\mathcal{T}_{\rm{LISA}}$ = 75 \% $\cdot$ 4 years \cite{Caprini:2019egz}  and $\mathcal{T}_{\rm{AEDGE}}$ = 3 years \cite{AEDGE:2019nxb}.
We will assume that a signal could be observed if $\mathrm{SNR}>10$, which is the usual criterion.

\begin{figure}[h!t]
\center
\includegraphics[width=.45\textwidth]{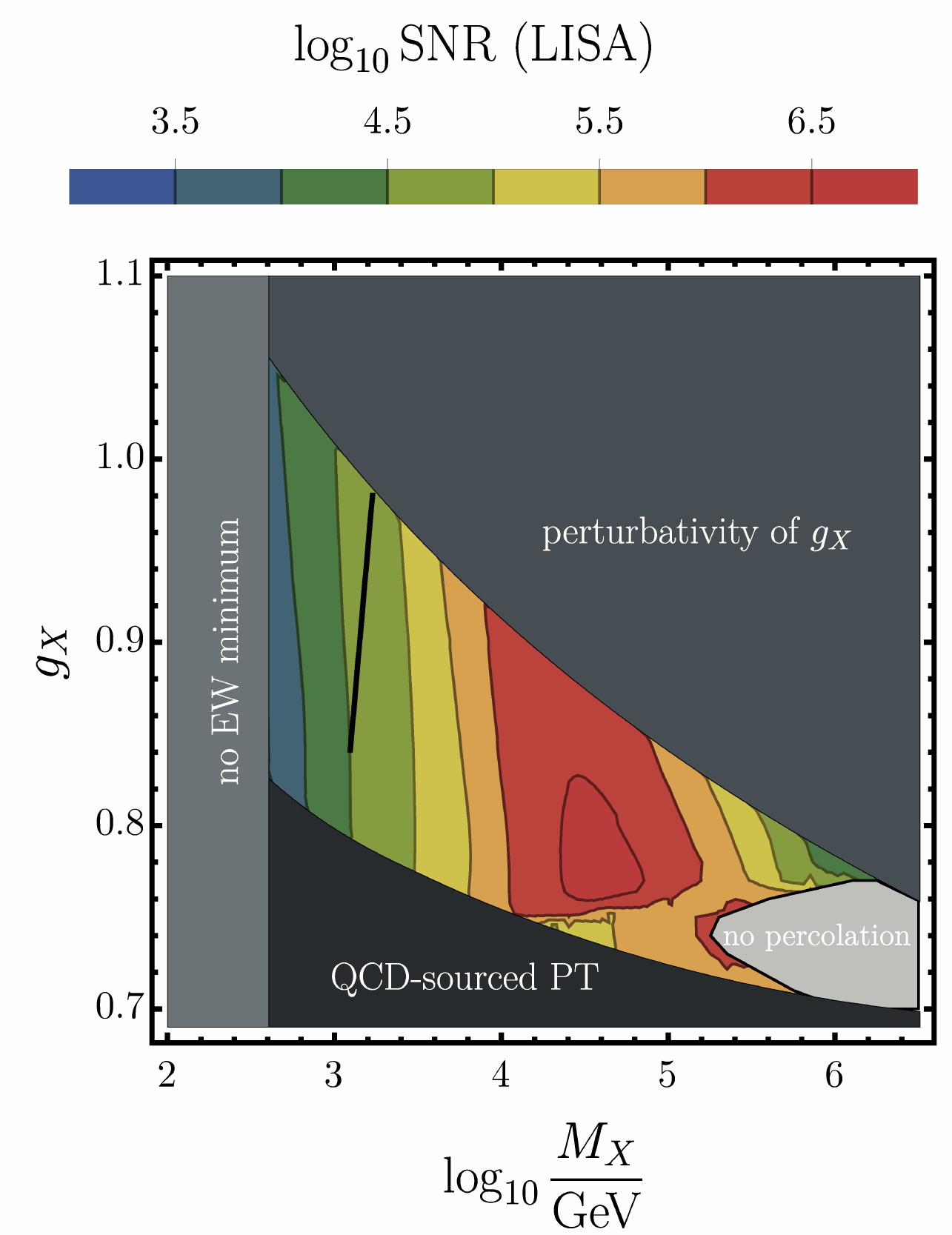}\hspace{20pt}
\includegraphics[width=.45\textwidth]{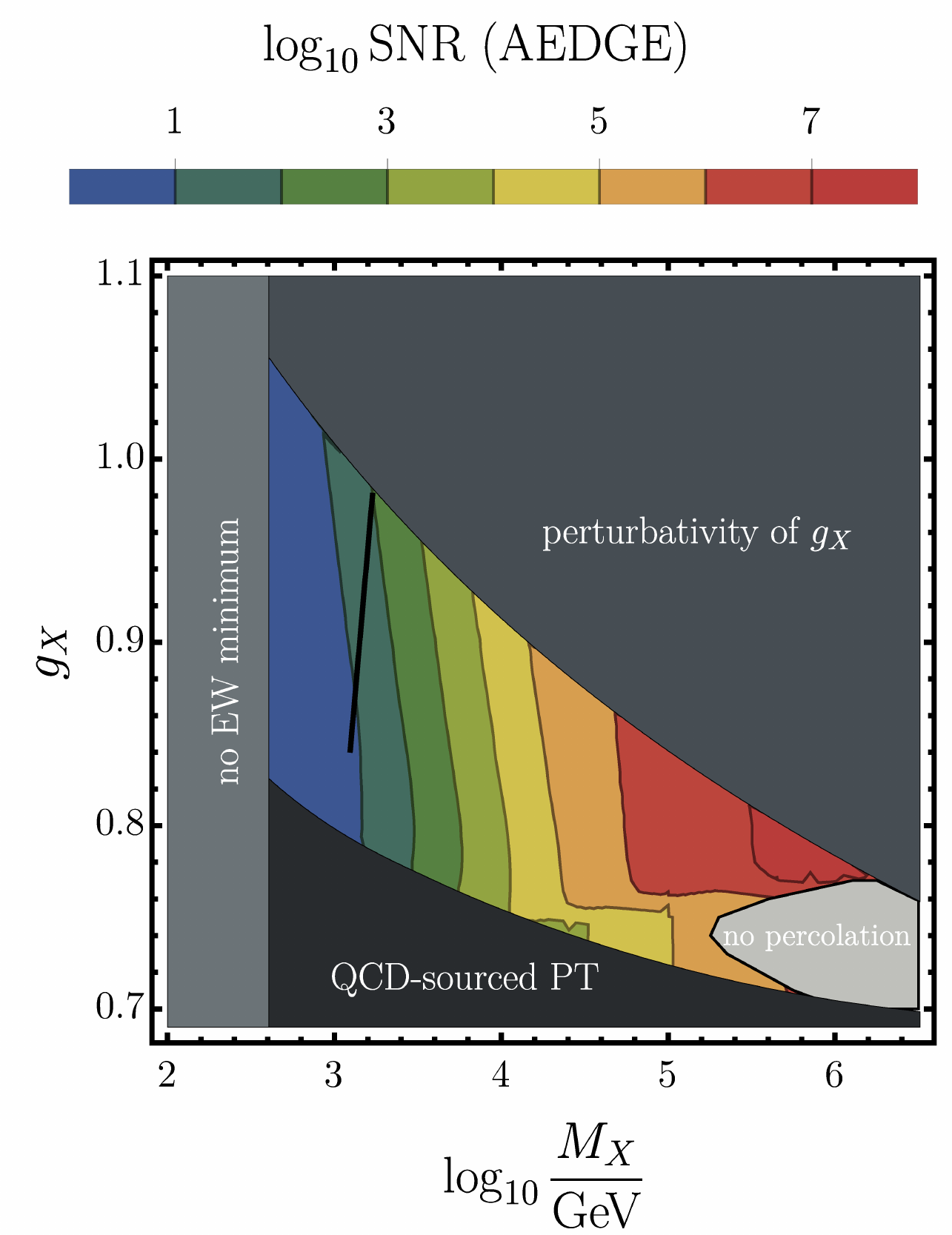}
\caption{Results for the signal-to-noise ratio for LISA (left panel) and AEDGE (right panel) for the predicted GW signal. The black line corresponds to the points that reproduce the measured DM relic abundance and also evade the DM direct detection experimental constraints.\label{fig:SNR}}
\end{figure}

The results are presented in figure~\ref{fig:SNR}. Superimposed is a curve indicating where in the parameter space the correct DM relic density is reproduced and the DM direct detection constraints are satisfied (solid black). Strikingly, the SNR for LISA for the predicted signal is above the observability threshold within the whole parameter space, and almost whole in the case of AEDGE. This means that a first-order phase transition sourced by tunnelling of a scalar field in the present model should be thoroughly testable by LISA and AEDGE. Moreover, in case of not observing a signal consistent with the expectations for the first-order phase transitions this scenario could be falsified.

\begin{figure}[h!t]
\center
\includegraphics[width=.8\textwidth]{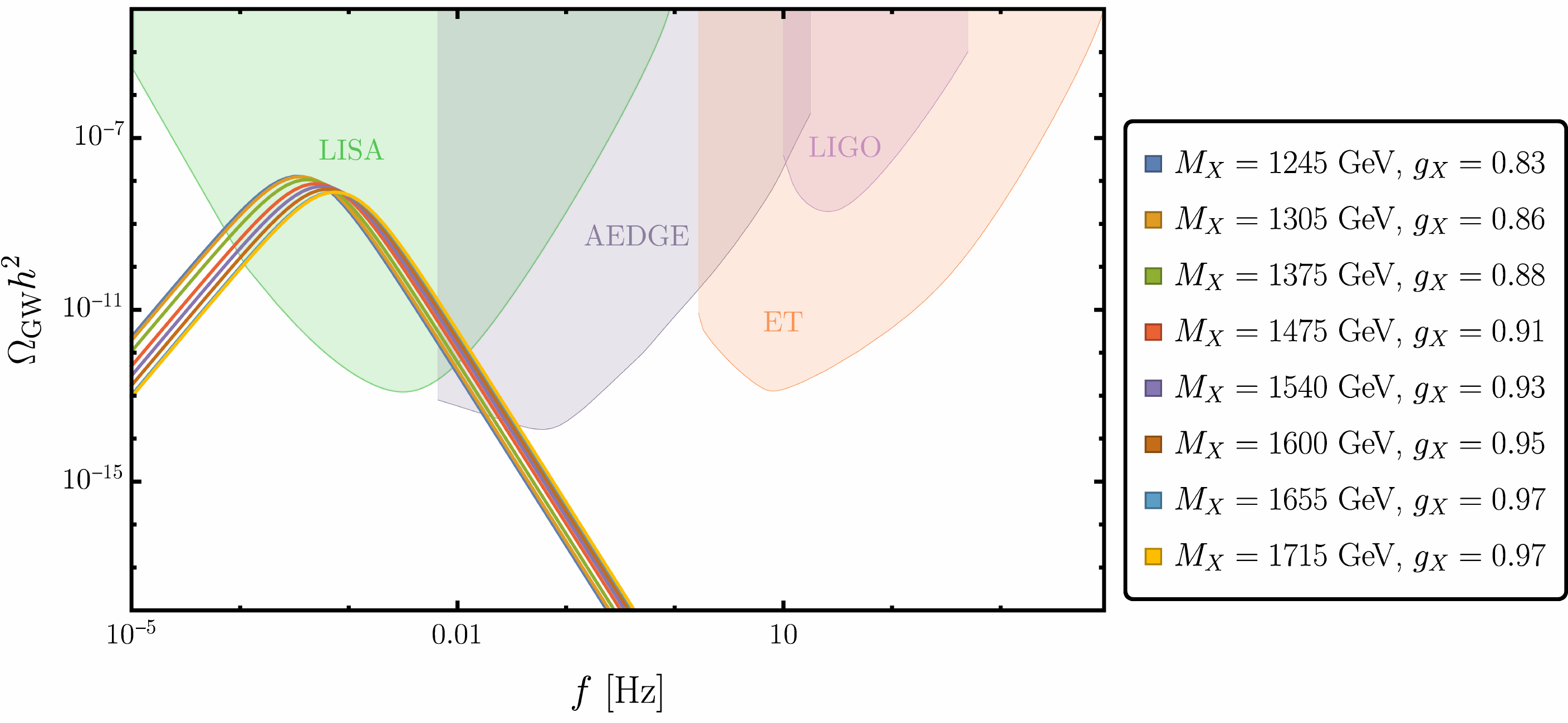}
\caption{Predictions for spectra of gravitational waves together with integrated sensitivity curves for LISA, AEDGE, ET and LIGO for the points in the parameter space where DM relic abundance is saturated.\label{fig:spectra-DM}}
\end{figure}

The correct DM relic abundance and non-exclusion by direct detection experiments (solid black line in figure~\ref{fig:SNR}) are located in the region of a relatively weaker signal. It is still well observable with LISA and AEDGE. %\comment{\sout{(this will depend on how good the sensitivity will be for lower frequencies)}}. 
The GW signal in the region where the correct abundance is reproduced is sourced entirely by sound waves. Examples of spectra for points along the black line in figure~\ref{fig:SNR} are shown in figure~\ref{fig:spectra-DM}.
%

%%%%%%%%%%%%%%%%%%%%%%%%%%%%%%%%%%%%%%%%%%%%%%%
\subsection{Renormalisation-scale dependence}
%%%%%%%%%%%%%%%%%%%%%%%%%%%%%%%%%%%%%%%%%%%%%%%

%
\begin{figure}[h!t]
\center
\includegraphics[width=.45\textwidth]{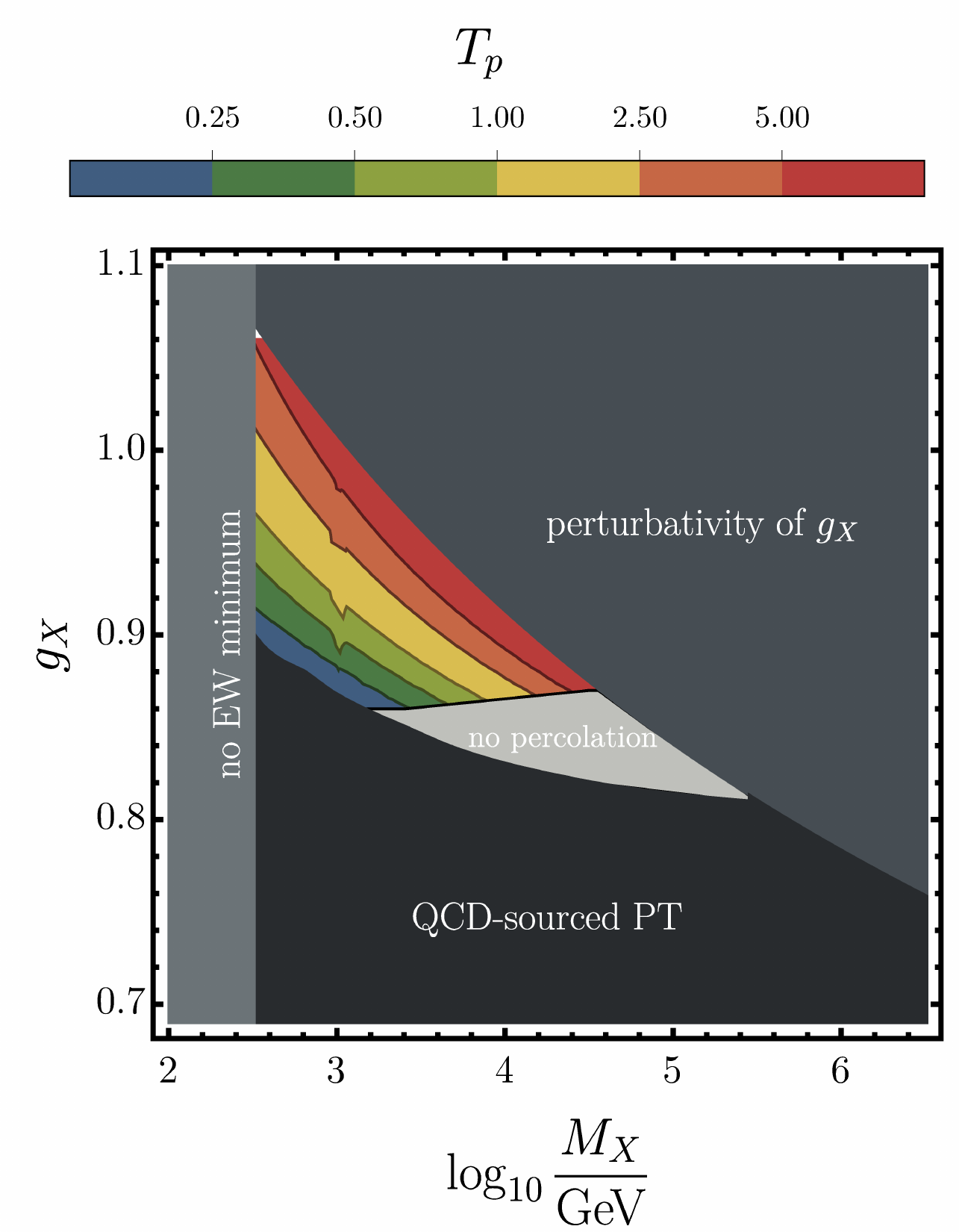}%\hspace{20pt}
\includegraphics[width=.45\textwidth]{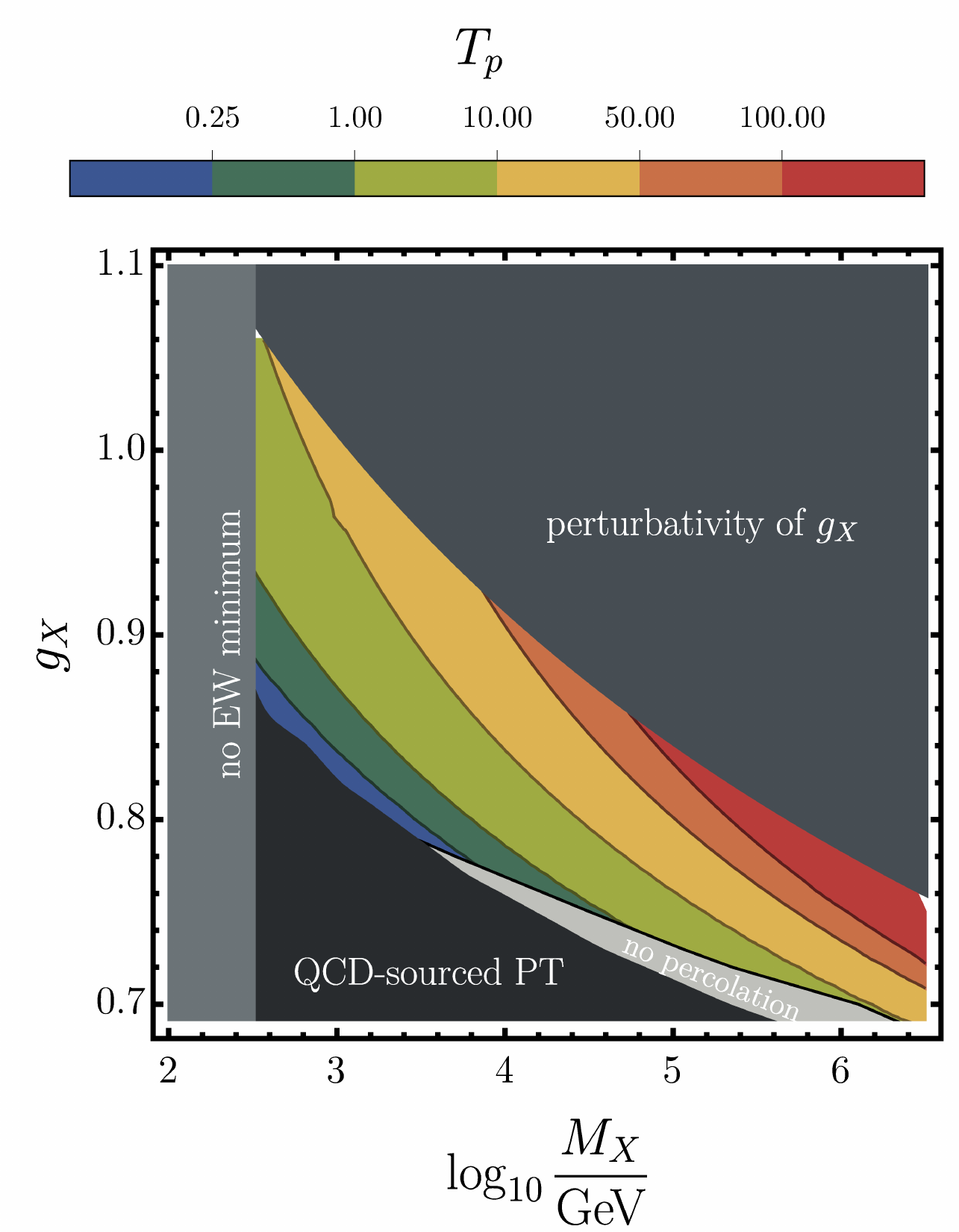}%\\
\caption{Results of the scan with fixed renormalisation scale, $\mu=M_X$ (left), $\mu=M_Z$ (right) for the percolation temperature $T_p$. %(upper row), $\log_{10}\alpha_*$ (lower row).
\label{fig:scan-at-fixed-mu}
}
\end{figure}

Finally, we perform scans of the parameter space at fixed $\mu$. This will tell us how our understanding of the parameter space and observability of the GW signal depends on the renormalisation scale. Figure~\ref{fig:scan-at-fixed-mu} shows the results for the percolation temperature $T_p$ computed at different scales ($\mu=M_X$ (left), $\mu=M_Z$ (right)) together with the previous constraints on the parameter space. Both figures indicate a striking dependence on the renormalisation scale. This has further implications since for $T_p\lesssim 0.1\g$ the PT is believed to be sourced by the QCD effects, which changes the nature and properties of the PT. In this work we focus on the PT sourced by the tunnelling, therefore the considered parameter space changes dramatically as the renormalisation scale is changed. Also, the answer to a basic question -- whether or not the PT completes via percolation of bubbles of the true vacuum -- is altered by the change of the renormalisation scale as can be seen by examining the percolation criterion (light-grey shaded region). 
%The strength of the transition is significantly modified as compared to the RG-improved case -- the strongest transitions with the largest $\alpha_*$ are absent and thus there are no valid points with GW sourced by bubble collisions. %We do not display $\kappa_{\textrm{sw}}$ as it is equal to 1 throughout the allowed parameter space.
The results show that the change of the scale at which computations are performed not only changes the results quantitatively, by shifting the values of the characteristic parameters of the phase transition, but it also significantly modifies them qualitatively -- by modifying the character of the phase transition, the very fact of its completion and the dominant source of the GW signal.

%%%%%%%%%%%%%%%%%%%%%%%%%%%%%%%%%%%%%%%%%%%%%%%
%%%%%%%%%%%%%%%%%%%%%%%%%%%%%%%%%%%%%%%%%%%%%%%
\section{Summary and conclusions}
%%%%%%%%%%%%%%%%%%%%%%%%%%%%%%%%%%%%%%%%%%%%%%%
%%%%%%%%%%%%%%%%%%%%%%%%%%%%%%%%%%%%%%%%%%%%%%%

In the present work, we studied a model endowed with classical scale invariance, a dark SU(2)$_X$ gauge group and a scalar doublet of this group. This model provides a dynamical mechanism of generating all the mass scales via radiative symmetry breaking, while featuring only two free parameters. Moreover, it provides dark matter candidates -- the three gauge bosons of the SU(2)$_X$ group which are degenerate in mass -- stabilised by an intrinsic $\mathbb{Z}_2 \times \mathbb{Z}_2'$ symmetry. Like other models with scaling symmetry, the studied model exhibits strong supercooling which results in the generation of an observable gravitational-wave signal.

Motivated by these attractive features we performed an analysis of the phase transition, gravitational wave generation and dark matter relic abundance, updating and extending the existing results~\cite{Hambye:2013, Carone:2013, Khoze:2014, Pelaggi:2014wba, Karam:2015, Plascencia:2016, Chataignier:2018, Hambye:2018, Baldes:2018, Prokopec:2018, Marfatia:2020}. The analysis features the key ingredients: 
\begin{itemize}
\item careful analysis of the potential in the light of radiative symmetry breaking; %in particular, a consistent expansion in the powers of couplings;
\item using renormalisation-group improved potential which includes all the leading order terms;
\item using RG-running to move between various relevant scales: the electroweak scale for scalar mass generation, the scale of the mass of the new scalar for its decay during reheating;
\item careful analysis of the supercooled phase transition, following recent developments, in particular imposing the percolation criterion which proved crucial for phenomenological predictions;
\item analysis of dark matter relic abundance in the light of the updated picture of the phase transition; %which excluded the supercool scenario (at least when the phase transition is sourced by the tunnelling, not QCD);
\item analysis of gravitational-wave spectra using most recent results from simulations;
\item using fixed-scale potential, in addition to the renormalisation-group-improved one, to study the scale dependence of the results.
\end{itemize}

The first and foremost result of our analysis is that within the model the gravitational wave signal sourced by a first-order phase transition associated with the $SU(2)_X$ and electroweak symmetry breaking is strong and observable for the whole allowed parameter space. This is an important conclusion since it allows this scenario to be falsified in case of negative LISA results. 

Second, we exclude the supercool dark matter scenario within the region where the phase transition proceeds via nucleation and percolation of bubbles of the true vacuum. It is a result of a combination of two reasons: we include the percolation condition, eq.~\eqref{eq:percolation-crit}, which allows to verify that a strongly supercooled phase transition indeed completes via percolation of bubbles and strongly constrains the parameter space relevant for our analysis. Moreover, we improve on the computation of the decay rate of the scalar field $\f$, which controls the reheating rate, which pushes the onset of inefficient reheating towards higher $M_X$, beyond the region of interest.

Third, we find the parameter space in which the correct relic dark matter abundance is predicted. It is produced via the standard freeze-out mechanism in the region with relatively low $M_X$ and large $g_X$. It is the region where the phase transition is relatively weak (compared with other regions of the parameter space), yet the gravitational-wave signal should be well observable with LISA. This parameter space is further reduced due to the recent direct detection constraints.

Moreover, in the present work we focused on the issue of scale dependence of the predictions. Our approach to reducing this dependence was to implement the renormalisation-group improvement procedure, respecting the power counting of couplings to include all the relevant terms. For comparison, we present results of computations performed at fixed scale, where the dependence on the renormalisation scale is significant. It is important to note that with the change of the scale the predictions do not only change quantitatively, they can change qualitatively. For example, for computations performed at a fixed scale (both $\mu=M_X$ and $M_Z$)  gravitational waves sourced by bubble collisions are not present. At the same time, with RG improvement we see a substantial region where bubble collisions are efficient in producing an observable signal.

%A possible issue to be studied further is to carefully verify the scaling of various couplings in the temperature-dependent setting, remembering that the high-temperature expansion is in general not valid in presence of radiative-symmetry breaking. Then one could check whether with the one-loop (and daisy-resummed) terms included in this work the scale dependence is indeed cancelled at the considered order %strictly preserved 
%or whether some higher-loop terms are needed. Another way of improving the accuracy of the predictions would be with the use of dimensional reduction, which is not straightforward in the case of radiative symmetry breaking. These issues will be a subject of a forthcoming work.

To sum up, the classically scale-invariant model with an extra SU(2) symmetry remains a valid theoretical framework for describing dark matter and gravitational-wave signal produced during a first-order phase transition in the early Universe. It will be tested experimentally by LISA and other gravitational-wave detectors. The predictions, however, are sensitive to the theoretical procedures implemented. Therefore, it is crucial to improve our understanding of theoretical pitfalls affecting the predictions. The present work is a step in this direction.

%-------------------------------------------------------------------------------
\acknowledgments{
%-------------------------------------------------------------------------------
We would like to thank Kristjan Kannike, Wojciech Kotlarski, Luca Marzola, Tania Robens, Martti Raidal and Rui Santos for useful discussions. We are indebted to Marek Lewicki for numerous discussions, clarifications and sharing data for the SNR plots. We are grateful to Jo\~{a}o Viana for his computation of Higgs decay width using \texttt{hdecay} and IT hints. We would also like to thank Matti Heikinheimo, Tomislav Prokopec, Tommi Tenkanen, Kimmo Tuominen and Ville Vaskonen for collaboration in the early stages of this work. AK was supported by the Estonian Research Council grants MOBTT5, MOBTT86, PSG761 and by the EU through the European Regional Development Fund CoE program TK133 ``The Dark Side of the Universe". The work of B\'{S} and MK is supported by the National Science Centre, Poland, through the SONATA project number 2018/31/D/ST2/03302.} AK would also like the thank the organizers of the ``School and Workshops on Elementary Particle Physics and Gravity", Corfu 2022, for their hospitality during his stay, and for giving him the opportunity to present this work.

\bibliography{conformal-bib+GW.bib}{}

\providecommand{\href}[2]{#2}\begingroup\raggedright\begin{thebibliography}{100}

\bibitem{Kierkla:2022odc}
M.~Kierkla, A.~Karam, and B.~Swiezewska, ``{Conformal model for gravitational
  waves and dark matter: a status update},''
  \href{http://dx.doi.org/10.1007/JHEP03(2023)007}{{\em JHEP} {\bfseries 03}
  (2023) 007}, \href{http://arxiv.org/abs/2210.07075}{{\ttfamily
  arXiv:2210.07075 [astro-ph.CO]}}.

\bibitem{Abbott:2016-2}
{\bfseries Virgo, LIGO Scientific} Collaboration, B.~P. Abbott {\em et~al.},
  ``{Observation of Gravitational Waves from a Binary Black Hole Merger},''
  \href{http://dx.doi.org/10.1103/PhysRevLett.116.061102}{{\em Phys. Rev.
  Lett.} {\bfseries 116} no.~6, (2016) 061102},
\href{http://arxiv.org/abs/1602.03837}{{\ttfamily arXiv:1602.03837 [gr-qc]}}.
%%CITATION = ARXIV:1602.03837;%%.

\bibitem{Abbott:2016}
{\bfseries Virgo, LIGO Scientific} Collaboration, B.~P. Abbott {\em et~al.},
  ``{GW151226: Observation of Gravitational Waves from a 22-Solar-Mass Binary
  Black Hole Coalescence},''
  \href{http://dx.doi.org/10.1103/PhysRevLett.116.241103}{{\em Phys. Rev.
  Lett.} {\bfseries 116} no.~24, (2016) 241103},
\href{http://arxiv.org/abs/1606.04855}{{\ttfamily arXiv:1606.04855 [gr-qc]}}.
%%CITATION = ARXIV:1606.04855;%%.

\bibitem{Abbott:2017}
{\bfseries VIRGO, LIGO Scientific} Collaboration, B.~P. Abbott {\em et~al.},
  ``{GW170104: Observation of a 50-Solar-Mass Binary Black Hole Coalescence at
  Redshift 0.2},'' \href{http://dx.doi.org/10.1103/PhysRevLett.118.221101,
  10.1103/PhysRevLett.121.129901}{{\em Phys. Rev. Lett.} {\bfseries 118}
  no.~22, (2017) 221101}, \href{http://arxiv.org/abs/1706.01812}{{\ttfamily
  arXiv:1706.01812 [gr-qc]}}.
[Erratum: Phys. Rev. Lett.121,no.12,129901(2018)].
%%CITATION = ARXIV:1706.01812;%%.

\bibitem{Abbott:2017-2}
{\bfseries Virgo, LIGO Scientific} Collaboration, B.~Abbott {\em et~al.},
  ``{GW170817: Observation of Gravitational Waves from a Binary Neutron Star
  Inspiral},'' \href{http://dx.doi.org/10.1103/PhysRevLett.119.161101}{{\em
  Phys. Rev. Lett.} {\bfseries 119} no.~16, (2017) 161101},
\href{http://arxiv.org/abs/1710.05832}{{\ttfamily arXiv:1710.05832 [gr-qc]}}.
%%CITATION = ARXIV:1710.05832;%%.

\bibitem{LIGOScientific:2017ycc}
{\bfseries LIGO Scientific, Virgo} Collaboration, B.~P. Abbott {\em et~al.},
  ``{GW170814: A Three-Detector Observation of Gravitational Waves from a
  Binary Black Hole Coalescence},''
  \href{http://dx.doi.org/10.1103/PhysRevLett.119.141101}{{\em Phys. Rev.
  Lett.} {\bfseries 119} no.~14, (2017) 141101},
  \href{http://arxiv.org/abs/1709.09660}{{\ttfamily arXiv:1709.09660 [gr-qc]}}.

\bibitem{LIGOScientific:2017vox}
{\bfseries LIGO Scientific, Virgo} Collaboration, B.~. P.~. Abbott {\em
  et~al.}, ``{GW170608: Observation of a 19-solar-mass Binary Black Hole
  Coalescence},'' \href{http://dx.doi.org/10.3847/2041-8213/aa9f0c}{{\em
  Astrophys. J. Lett.} {\bfseries 851} (2017) L35},
  \href{http://arxiv.org/abs/1711.05578}{{\ttfamily arXiv:1711.05578
  [astro-ph.HE]}}.

\bibitem{Bartolo:2016ami}
N.~Bartolo {\em et~al.}, ``{Science with the space-based interferometer LISA.
  IV: Probing inflation with gravitational waves},''
  \href{http://dx.doi.org/10.1088/1475-7516/2016/12/026}{{\em JCAP} {\bfseries
  12} (2016) 026}, \href{http://arxiv.org/abs/1610.06481}{{\ttfamily
  arXiv:1610.06481 [astro-ph.CO]}}.

\bibitem{Caprini:2019pxz}
C.~Caprini, D.~G. Figueroa, R.~Flauger, G.~Nardini, M.~Peloso, M.~Pieroni,
  A.~Ricciardone, and G.~Tasinato, ``{Reconstructing the spectral shape of a
  stochastic gravitational wave background with LISA},''
  \href{http://dx.doi.org/10.1088/1475-7516/2019/11/017}{{\em JCAP} {\bfseries
  11} (2019) 017}, \href{http://arxiv.org/abs/1906.09244}{{\ttfamily
  arXiv:1906.09244 [astro-ph.CO]}}.

\bibitem{Gowling:2021gcy}
C.~Gowling and M.~Hindmarsh, ``{Observational prospects for phase transitions
  at LISA: Fisher matrix analysis},''
  \href{http://dx.doi.org/10.1088/1475-7516/2021/10/039}{{\em JCAP} {\bfseries
  10} (2021) 039}, \href{http://arxiv.org/abs/2106.05984}{{\ttfamily
  arXiv:2106.05984 [astro-ph.CO]}}.

\bibitem{LISACosWG:2022jok}
{\bfseries LISA Cosmology Working Group} Collaboration, P.~Auclair {\em
  et~al.}, ``{Cosmology with the Laser Interferometer Space Antenna},''
  \href{http://arxiv.org/abs/2204.05434}{{\ttfamily arXiv:2204.05434
  [astro-ph.CO]}}.

\bibitem{Boileau:2022ter}
G.~Boileau, N.~Christensen, C.~Gowling, M.~Hindmarsh, and R.~Meyer,
  ``{Prospects for LISA to detect a gravitational-wave background from first
  order phase transitions},'' \href{http://arxiv.org/abs/2209.13277}{{\ttfamily
  arXiv:2209.13277 [gr-qc]}}.

\bibitem{Gowling:2022pzb}
C.~Gowling, M.~Hindmarsh, D.~C. Hooper, and J.~Torrado, ``{Reconstructing
  physical parameters from template gravitational wave spectra at LISA: first
  order phase transitions},'' \href{http://arxiv.org/abs/2209.13551}{{\ttfamily
  arXiv:2209.13551 [astro-ph.CO]}}.

\bibitem{Badurina:2019hst}
L.~Badurina {\em et~al.}, ``{AION: An Atom Interferometer Observatory and
  Network},'' \href{http://dx.doi.org/10.1088/1475-7516/2020/05/011}{{\em JCAP}
  {\bfseries 05} (2020) 011}, \href{http://arxiv.org/abs/1911.11755}{{\ttfamily
  arXiv:1911.11755 [astro-ph.CO]}}.

\bibitem{Graham:2016plp}
P.~W. Graham, J.~M. Hogan, M.~A. Kasevich, and S.~Rajendran, ``{Resonant mode
  for gravitational wave detectors based on atom interferometry},''
  \href{http://dx.doi.org/10.1103/PhysRevD.94.104022}{{\em Phys. Rev. D}
  {\bfseries 94} no.~10, (2016) 104022},
  \href{http://arxiv.org/abs/1606.01860}{{\ttfamily arXiv:1606.01860
  [physics.atom-ph]}}.

\bibitem{Graham:2017pmn}
{\bfseries MAGIS} Collaboration, P.~W. Graham, J.~M. Hogan, M.~A. Kasevich,
  S.~Rajendran, and R.~W. Romani, ``{Mid-band gravitational wave detection with
  precision atomic sensors},''
  \href{http://arxiv.org/abs/1711.02225}{{\ttfamily arXiv:1711.02225
  [astro-ph.IM]}}.

\bibitem{AEDGE:2019nxb}
{\bfseries AEDGE} Collaboration, Y.~A. El-Neaj {\em et~al.}, ``{AEDGE: Atomic
  Experiment for Dark Matter and Gravity Exploration in Space},''
  \href{http://dx.doi.org/10.1140/epjqt/s40507-020-0080-0}{{\em EPJ Quant.
  Technol.} {\bfseries 7} (2020) 6},
  \href{http://arxiv.org/abs/1908.00802}{{\ttfamily arXiv:1908.00802 [gr-qc]}}.

\bibitem{Punturo:2010zz}
M.~Punturo {\em et~al.}, ``{The Einstein Telescope: A third-generation
  gravitational wave observatory},''
  \href{http://dx.doi.org/10.1088/0264-9381/27/19/194002}{{\em Class. Quant.
  Grav.} {\bfseries 27} (2010) 194002}.

\bibitem{Hild:2010id}
S.~Hild {\em et~al.}, ``{Sensitivity Studies for Third-Generation Gravitational
  Wave Observatories},''
  \href{http://dx.doi.org/10.1088/0264-9381/28/9/094013}{{\em Class. Quant.
  Grav.} {\bfseries 28} (2011) 094013},
  \href{http://arxiv.org/abs/1012.0908}{{\ttfamily arXiv:1012.0908 [gr-qc]}}.

\bibitem{Harry:2010zz}
{\bfseries LIGO Scientific} Collaboration, G.~M. Harry, ``{Advanced LIGO: The
  next generation of gravitational wave detectors},''
  \href{http://dx.doi.org/10.1088/0264-9381/27/8/084006}{{\em Class. Quant.
  Grav.} {\bfseries 27} (2010) 084006}.

\bibitem{VIRGO:2014yos}
{\bfseries VIRGO} Collaboration, F.~Acernese {\em et~al.}, ``{Advanced Virgo: a
  second-generation interferometric gravitational wave detector},''
  \href{http://dx.doi.org/10.1088/0264-9381/32/2/024001}{{\em Class. Quant.
  Grav.} {\bfseries 32} no.~2, (2015) 024001},
  \href{http://arxiv.org/abs/1408.3978}{{\ttfamily arXiv:1408.3978 [gr-qc]}}.

\bibitem{LIGOScientific:2014pky}
{\bfseries LIGO Scientific} Collaboration, J.~Aasi {\em et~al.}, ``{Advanced
  LIGO},'' \href{http://dx.doi.org/10.1088/0264-9381/32/7/074001}{{\em Class.
  Quant. Grav.} {\bfseries 32} (2015) 074001},
  \href{http://arxiv.org/abs/1411.4547}{{\ttfamily arXiv:1411.4547 [gr-qc]}}.

\bibitem{LIGOScientific:2019lzm}
{\bfseries LIGO Scientific, Virgo} Collaboration, R.~Abbott {\em et~al.},
  ``{Open data from the first and second observing runs of Advanced LIGO and
  Advanced Virgo},'' \href{http://dx.doi.org/10.1016/j.softx.2021.100658}{{\em
  SoftwareX} {\bfseries 13} (2021) 100658},
  \href{http://arxiv.org/abs/1912.11716}{{\ttfamily arXiv:1912.11716 [gr-qc]}}.

\bibitem{Caprini:2015}
C.~Caprini {\em et~al.}, ``{Science with the space-based interferometer eLISA.
  II: Gravitational waves from cosmological phase transitions},''
  \href{http://dx.doi.org/10.1088/1475-7516/2016/04/001}{{\em JCAP} {\bfseries
  1604} no.~04, (2016) 001},
\href{http://arxiv.org/abs/1512.06239}{{\ttfamily arXiv:1512.06239
  [astro-ph.CO]}}.
%%CITATION = ARXIV:1512.06239;%%.

\bibitem{Hambye:2013}
T.~Hambye and A.~Strumia, ``{Dynamical generation of the weak and Dark Matter
  scale},'' \href{http://dx.doi.org/10.1103/PhysRevD.88.055022}{{\em Phys.
  Rev.} {\bfseries D88} (2013) 055022},
\href{http://arxiv.org/abs/1306.2329}{{\ttfamily arXiv:1306.2329 [hep-ph]}}.
%%CITATION = ARXIV:1306.2329;%%.

\bibitem{Jaeckel:2016}
J.~Jaeckel, V.~V. Khoze, and M.~Spannowsky, ``{Hearing the signal of dark
  sectors with gravitational wave detectors},''
  \href{http://dx.doi.org/10.1103/PhysRevD.94.103519}{{\em Phys. Rev.}
  {\bfseries D94} no.~10, (2016) 103519},
\href{http://arxiv.org/abs/1602.03901}{{\ttfamily arXiv:1602.03901 [hep-ph]}}.
%%CITATION = ARXIV:1602.03901;%%.

\bibitem{Hashino:2016}
K.~Hashino, M.~Kakizaki, S.~Kanemura, and T.~Matsui, ``{Synergy between
  measurements of gravitational waves and the triple-Higgs coupling in probing
  the first-order electroweak phase transition},''
  \href{http://dx.doi.org/10.1103/PhysRevD.94.015005}{{\em Phys. Rev.}
  {\bfseries D94} no.~1, (2016) 015005},
\href{http://arxiv.org/abs/1604.02069}{{\ttfamily arXiv:1604.02069 [hep-ph]}}.
%%CITATION = ARXIV:1604.02069;%%.

\bibitem{Jinno:2016}
R.~Jinno and M.~Takimoto, ``{Probing a classically conformal B-L model with
  gravitational waves},''
  \href{http://dx.doi.org/10.1103/PhysRevD.95.015020}{{\em Phys. Rev.}
  {\bfseries D95} no.~1, (2017) 015020},
\href{http://arxiv.org/abs/1604.05035}{{\ttfamily arXiv:1604.05035 [hep-ph]}}.
%%CITATION = ARXIV:1604.05035;%%.

\bibitem{Marzola:2017}
L.~Marzola, A.~Racioppi, and V.~Vaskonen, ``{Phase transition and gravitational
  wave phenomenology of scalar conformal extensions of the Standard Model},''
  \href{http://dx.doi.org/10.1140/epjc/s10052-017-4996-1}{{\em Eur. Phys. J.}
  {\bfseries C77} no.~7, (2017) 484},
\href{http://arxiv.org/abs/1704.01034}{{\ttfamily arXiv:1704.01034 [hep-ph]}}.
%%CITATION = ARXIV:1704.01034;%%.

\bibitem{Ghorbani:2017lyk}
P.~H. Ghorbani, ``{Electroweak phase transition in the scale invariant standard
  model},'' \href{http://dx.doi.org/10.1103/PhysRevD.98.115016}{{\em Phys. Rev.
  D} {\bfseries 98} no.~11, (2018) 115016},
  \href{http://arxiv.org/abs/1711.11541}{{\ttfamily arXiv:1711.11541
  [hep-ph]}}.

\bibitem{Baldes:2018}
I.~Baldes and C.~Garcia-Cely, ``{Strong gravitational radiation from a simple
  dark matter model},'' \href{http://dx.doi.org/10.1007/JHEP05(2019)190}{{\em
  JHEP} {\bfseries 05} (2019) 190},
  \href{http://arxiv.org/abs/1809.01198}{{\ttfamily arXiv:1809.01198
  [hep-ph]}}.

\bibitem{Prokopec:2018}
T.~Prokopec, J.~Rezacek, and B.~\'Swie\.zewska, ``{Gravitational waves from
  conformal symmetry breaking},''
  \href{http://dx.doi.org/10.1088/1475-7516/2019/02/009}{{\em JCAP} {\bfseries
  02} (2019) 009}, \href{http://arxiv.org/abs/1809.11129}{{\ttfamily
  arXiv:1809.11129 [hep-ph]}}.

\bibitem{Marzo:2018}
C.~Marzo, L.~Marzola, and V.~Vaskonen, ``{Phase transition and vacuum stability
  in the classically conformal B\textendash{}L model},''
  \href{http://dx.doi.org/10.1140/epjc/s10052-019-7076-x}{{\em Eur. Phys. J. C}
  {\bfseries 79} no.~7, (2019) 601},
  \href{http://arxiv.org/abs/1811.11169}{{\ttfamily arXiv:1811.11169
  [hep-ph]}}.

\bibitem{Mohamadnejad:2019vzg}
A.~Mohamadnejad, ``{Gravitational waves from scale-invariant vector dark matter
  model: Probing below the neutrino-floor},''
  \href{http://dx.doi.org/10.1140/epjc/s10052-020-7756-6}{{\em Eur. Phys. J. C}
  {\bfseries 80} no.~3, (2020) 197},
  \href{http://arxiv.org/abs/1907.08899}{{\ttfamily arXiv:1907.08899
  [hep-ph]}}.

\bibitem{Ghoshal:2020vud}
A.~Ghoshal and A.~Salvio, ``{Gravitational waves from fundamental axion
  dynamics},'' \href{http://dx.doi.org/10.1007/JHEP12(2020)049}{{\em JHEP}
  {\bfseries 12} (2020) 049}, \href{http://arxiv.org/abs/2007.00005}{{\ttfamily
  arXiv:2007.00005 [hep-ph]}}.

\bibitem{Kang:2020jeg}
Z.~Kang and J.~Zhu, ``{Scale-genesis by Dark Matter and Its Gravitational Wave
  Signal},'' \href{http://dx.doi.org/10.1103/PhysRevD.102.053011}{{\em Phys.
  Rev. D} {\bfseries 102} no.~5, (2020) 053011},
  \href{http://arxiv.org/abs/2003.02465}{{\ttfamily arXiv:2003.02465
  [hep-ph]}}.

\bibitem{Mohamadnejad:2021tke}
A.~Mohamadnejad, ``{Electroweak phase transition and gravitational waves in a
  two-component dark matter model},''
  \href{http://dx.doi.org/10.1007/JHEP03(2022)188}{{\em JHEP} {\bfseries 03}
  (2022) 188}, \href{http://arxiv.org/abs/2111.04342}{{\ttfamily
  arXiv:2111.04342 [hep-ph]}}.

\bibitem{Dasgupta:2022isg}
A.~Dasgupta, P.~S.~B. Dev, A.~Ghoshal, and A.~Mazumdar, ``{Gravitational Wave
  Pathway to Testable Leptogenesis},''
  \href{http://arxiv.org/abs/2206.07032}{{\ttfamily arXiv:2206.07032
  [hep-ph]}}.

\bibitem{Hempfling:1996}
R.~Hempfling, ``{The Next-to-minimal Coleman-Weinberg model},''
  \href{http://dx.doi.org/10.1016/0370-2693(96)00446-7}{{\em Phys. Lett.}
  {\bfseries B379} (1996) 153--158},
\href{http://arxiv.org/abs/hep-ph/9604278}{{\ttfamily arXiv:hep-ph/9604278
  [hep-ph]}}.
%%CITATION = HEP-PH/9604278;%%.

\bibitem{Sher:1996ib}
M.~Sher, ``{The Coleman-Weinberg phase transition in extended Higgs models},''
  \href{http://dx.doi.org/10.1103/PhysRevD.54.7071}{{\em Phys. Rev. D}
  {\bfseries 54} (1996) 7071--7074},
  \href{http://arxiv.org/abs/hep-ph/9607337}{{\ttfamily arXiv:hep-ph/9607337}}.

\bibitem{Chang:2007}
W.-F. Chang, J.~N. Ng, and J.~M.~S. Wu, ``{Shadow Higgs from a scale-invariant
  hidden U(1)(s) model},''
  \href{http://dx.doi.org/10.1103/PhysRevD.75.115016}{{\em Phys. Rev.}
  {\bfseries D75} (2007) 115016},
\href{http://arxiv.org/abs/hep-ph/0701254}{{\ttfamily arXiv:hep-ph/0701254
  [HEP-PH]}}.
%%CITATION = HEP-PH/0701254;%%.

\bibitem{Iso:2009}
S.~Iso, N.~Okada, and Y.~Orikasa, ``{Classically conformal $B-L$ extended
  Standard Model},''
  \href{http://dx.doi.org/10.1016/j.physletb.2009.04.046}{{\em Phys. Lett.}
  {\bfseries B676} (2009) 81--87},
\href{http://arxiv.org/abs/0902.4050}{{\ttfamily arXiv:0902.4050 [hep-ph]}}.
%%CITATION = ARXIV:0902.4050;%%.

\bibitem{Iso:2012jn}
S.~Iso and Y.~Orikasa, ``{TeV Scale B-L model with a flat Higgs potential at
  the Planck scale: In view of the hierarchy problem},''
  \href{http://dx.doi.org/10.1093/ptep/pts099}{{\em PTEP} {\bfseries 2013}
  (2013) 023B08}, \href{http://arxiv.org/abs/1210.2848}{{\ttfamily
  arXiv:1210.2848 [hep-ph]}}.

\bibitem{Khoze:2013-1}
C.~Englert, J.~Jaeckel, V.~V. Khoze, and M.~Spannowsky, ``{Emergence of the
  Electroweak Scale through the Higgs Portal},''
  \href{http://dx.doi.org/10.1007/JHEP04(2013)060}{{\em JHEP} {\bfseries 04}
  (2013) 060},
\href{http://arxiv.org/abs/1301.4224}{{\ttfamily arXiv:1301.4224 [hep-ph]}}.
%%CITATION = ARXIV:1301.4224;%%.

\bibitem{Khoze:2013-2}
V.~V. Khoze and G.~Ro, ``{Leptogenesis and Neutrino Oscillations in the
  Classically Conformal Standard Model with the Higgs Portal},''
  \href{http://dx.doi.org/10.1007/JHEP10(2013)075}{{\em JHEP} {\bfseries 10}
  (2013) 075},
\href{http://arxiv.org/abs/1307.3764}{{\ttfamily arXiv:1307.3764 [hep-ph]}}.
%%CITATION = ARXIV:1307.3764;%%.

\bibitem{Khoze:2013-3}
V.~V. Khoze, ``{Inflation and Dark Matter in the Higgs Portal of Classically
  Scale Invariant Standard Model},''
  \href{http://dx.doi.org/10.1007/JHEP11(2013)215}{{\em JHEP} {\bfseries 11}
  (2013) 215},
\href{http://arxiv.org/abs/1308.6338}{{\ttfamily arXiv:1308.6338 [hep-ph]}}.
%%CITATION = ARXIV:1308.6338;%%.

\bibitem{Hashimoto:2013hta}
M.~Hashimoto, S.~Iso, and Y.~Orikasa, ``{Radiative symmetry breaking at the
  Fermi scale and flat potential at the Planck scale},''
  \href{http://dx.doi.org/10.1103/PhysRevD.89.016019}{{\em Phys. Rev. D}
  {\bfseries 89} no.~1, (2014) 016019},
  \href{http://arxiv.org/abs/1310.4304}{{\ttfamily arXiv:1310.4304 [hep-ph]}}.

\bibitem{Hashimoto:2014ela}
M.~Hashimoto, S.~Iso, and Y.~Orikasa, ``{Radiative symmetry breaking from flat
  potential in various U(1)' models},''
  \href{http://dx.doi.org/10.1103/PhysRevD.89.056010}{{\em Phys. Rev. D}
  {\bfseries 89} no.~5, (2014) 056010},
  \href{http://arxiv.org/abs/1401.5944}{{\ttfamily arXiv:1401.5944 [hep-ph]}}.

\bibitem{Benic:2014xho}
S.~Benic and B.~Radovcic, ``{Electroweak breaking and Dark Matter from the
  common scale},'' \href{http://dx.doi.org/10.1016/j.physletb.2014.03.018}{{\em
  Phys. Lett. B} {\bfseries 732} (2014) 91--94},
  \href{http://arxiv.org/abs/1401.8183}{{\ttfamily arXiv:1401.8183 [hep-ph]}}.

\bibitem{Khoze:2014}
V.~V. Khoze, C.~McCabe, and G.~Ro, ``{Higgs vacuum stability from the dark
  matter portal},'' \href{http://dx.doi.org/10.1007/JHEP08(2014)026}{{\em JHEP}
  {\bfseries 08} (2014) 026},
\href{http://arxiv.org/abs/1403.4953}{{\ttfamily arXiv:1403.4953 [hep-ph]}}.
%%CITATION = ARXIV:1403.4953;%%.

\bibitem{Benic:2014aga}
S.~Benic and B.~Radovcic, ``{Majorana dark matter in a classically scale
  invariant model},'' \href{http://dx.doi.org/10.1007/JHEP01(2015)143}{{\em
  JHEP} {\bfseries 01} (2015) 143},
  \href{http://arxiv.org/abs/1409.5776}{{\ttfamily arXiv:1409.5776 [hep-ph]}}.

\bibitem{Okada:2014nea}
H.~Okada and Y.~Orikasa, ``{Classically conformal radiative neutrino model with
  gauged B \ensuremath{-} L symmetry},''
  \href{http://dx.doi.org/10.1016/j.physletb.2016.07.039}{{\em Phys. Lett. B}
  {\bfseries 760} (2016) 558--564},
  \href{http://arxiv.org/abs/1412.3616}{{\ttfamily arXiv:1412.3616 [hep-ph]}}.

\bibitem{Guo:2015}
J.~Guo, Z.~Kang, P.~Ko, and Y.~Orikasa, ``{Accidental dark matter: Case in the
  scale invariant local B-L model},''
  \href{http://dx.doi.org/10.1103/PhysRevD.91.115017}{{\em Phys. Rev.}
  {\bfseries D91} no.~11, (2015) 115017},
\href{http://arxiv.org/abs/1502.00508}{{\ttfamily arXiv:1502.00508 [hep-ph]}}.
%%CITATION = ARXIV:1502.00508;%%.

\bibitem{Humbert:2015epa}
P.~Humbert, M.~Lindner, and J.~Smirnov, ``{The Inverse Seesaw in Conformal
  Electro-Weak Symmetry Breaking and Phenomenological Consequences},''
  \href{http://dx.doi.org/10.1007/JHEP06(2015)035}{{\em JHEP} {\bfseries 06}
  (2015) 035}, \href{http://arxiv.org/abs/1503.03066}{{\ttfamily
  arXiv:1503.03066 [hep-ph]}}.

\bibitem{Oda:2015gna}
S.~Oda, N.~Okada, and D.-s. Takahashi, ``{Classically conformal U(1)' extended
  standard model and Higgs vacuum stability},''
  \href{http://dx.doi.org/10.1103/PhysRevD.92.015026}{{\em Phys. Rev. D}
  {\bfseries 92} no.~1, (2015) 015026},
  \href{http://arxiv.org/abs/1504.06291}{{\ttfamily arXiv:1504.06291
  [hep-ph]}}.

\bibitem{Humbert:2015yva}
P.~Humbert, M.~Lindner, S.~Patra, and J.~Smirnov, ``{Lepton Number Violation
  within the Conformal Inverse Seesaw},''
  \href{http://dx.doi.org/10.1007/JHEP09(2015)064}{{\em JHEP} {\bfseries 09}
  (2015) 064}, \href{http://arxiv.org/abs/1505.07453}{{\ttfamily
  arXiv:1505.07453 [hep-ph]}}.

\bibitem{Plascencia:2015}
A.~D. Plascencia, ``{Classical scale invariance in the inert doublet model},''
  \href{http://dx.doi.org/10.1007/JHEP09(2015)026}{{\em JHEP} {\bfseries 09}
  (2015) 026},
\href{http://arxiv.org/abs/1507.04996}{{\ttfamily arXiv:1507.04996 [hep-ph]}}.
%%CITATION = ARXIV:1507.04996;%%.

\bibitem{Haba:2015lka}
N.~Haba, H.~Ishida, N.~Okada, and Y.~Yamaguchi, ``{Bosonic seesaw mechanism in
  a classically conformal extension of the Standard Model},''
  \href{http://dx.doi.org/10.1016/j.physletb.2016.01.050}{{\em Phys. Lett. B}
  {\bfseries 754} (2016) 349--352},
  \href{http://arxiv.org/abs/1508.06828}{{\ttfamily arXiv:1508.06828
  [hep-ph]}}.

\bibitem{Das:2015nwk}
A.~Das, N.~Okada, and N.~Papapietro, ``{Electroweak vacuum stability in
  classically conformal B-L extension of the Standard Model},''
  \href{http://dx.doi.org/10.1140/epjc/s10052-017-4683-2}{{\em Eur. Phys. J. C}
  {\bfseries 77} no.~2, (2017) 122},
  \href{http://arxiv.org/abs/1509.01466}{{\ttfamily arXiv:1509.01466
  [hep-ph]}}.

\bibitem{Haba:2015nwl}
N.~Haba, H.~Ishida, R.~Takahashi, and Y.~Yamaguchi, ``{Gauge coupling
  unification in a classically scale invariant model},''
  \href{http://dx.doi.org/10.1007/JHEP02(2016)058}{{\em JHEP} {\bfseries 02}
  (2016) 058}, \href{http://arxiv.org/abs/1511.02107}{{\ttfamily
  arXiv:1511.02107 [hep-ph]}}.

\bibitem{Wang:2015sxe}
Z.-W. Wang, F.~S. Sage, T.~G. Steele, and R.~B. Mann, ``{Asymptotic Safety in
  the Conformal Hidden Sector?},''
  \href{http://dx.doi.org/10.1088/1361-6471/aad2c7}{{\em J. Phys. G} {\bfseries
  45} no.~9, (2018) 095002}, \href{http://arxiv.org/abs/1511.02531}{{\ttfamily
  arXiv:1511.02531 [hep-ph]}}.

\bibitem{Das:2016zue}
A.~Das, S.~Oda, N.~Okada, and D.-s. Takahashi, ``{Classically conformal U(1)'
  extended standard model, electroweak vacuum stability, and LHC Run-2
  bounds},'' \href{http://dx.doi.org/10.1103/PhysRevD.93.115038}{{\em Phys.
  Rev. D} {\bfseries 93} no.~11, (2016) 115038},
  \href{http://arxiv.org/abs/1605.01157}{{\ttfamily arXiv:1605.01157
  [hep-ph]}}.

\bibitem{Oda:2017kwl}
S.~Oda, N.~Okada, and D.-s. Takahashi, ``{Right-handed neutrino dark matter in
  the classically conformal U(1)' extended standard model},''
  \href{http://dx.doi.org/10.1103/PhysRevD.96.095032}{{\em Phys. Rev. D}
  {\bfseries 96} no.~9, (2017) 095032},
  \href{http://arxiv.org/abs/1704.05023}{{\ttfamily arXiv:1704.05023
  [hep-ph]}}.

\bibitem{Hambye:2018}
T.~Hambye, A.~Strumia, and D.~Teresi, ``{Super-cool Dark Matter},''
  \href{http://dx.doi.org/10.1007/JHEP08(2018)188}{{\em JHEP} {\bfseries 08}
  (2018) 188}, \href{http://arxiv.org/abs/1805.01473}{{\ttfamily
  arXiv:1805.01473 [hep-ph]}}.

\bibitem{Loebbert:2018}
F.~Loebbert, J.~Miczajka, and J.~Plefka, ``{Consistent Conformal Extensions of
  the Standard Model},''
  \href{http://dx.doi.org/10.1103/PhysRevD.99.015026}{{\em Phys. Rev. D}
  {\bfseries 99} no.~1, (2019) 015026},
  \href{http://arxiv.org/abs/1805.09727}{{\ttfamily arXiv:1805.09727
  [hep-ph]}}.

\bibitem{YaserAyazi:2019caf}
S.~Yaser~Ayazi and A.~Mohamadnejad, ``{Conformal vector dark matter and
  strongly first-order electroweak phase transition},''
  \href{http://dx.doi.org/10.1007/JHEP03(2019)181}{{\em JHEP} {\bfseries 03}
  (2019) 181}, \href{http://arxiv.org/abs/1901.04168}{{\ttfamily
  arXiv:1901.04168 [hep-ph]}}.

\bibitem{Kim:2019ogz}
Y.~G. Kim, K.~Y. Lee, and S.-H. Nam, ``{Conformal invariance and singlet
  fermionic dark matter},''
  \href{http://dx.doi.org/10.1103/PhysRevD.100.075038}{{\em Phys. Rev. D}
  {\bfseries 100} no.~7, (2019) 075038},
  \href{http://arxiv.org/abs/1906.03390}{{\ttfamily arXiv:1906.03390
  [hep-ph]}}.

\bibitem{Gialamas:2021enw}
I.~D. Gialamas, A.~Karam, T.~D. Pappas, and V.~C. Spanos, ``{Scale-invariant
  quadratic gravity and inflation in the Palatini formalism},''
  \href{http://dx.doi.org/10.1103/PhysRevD.104.023521}{{\em Phys. Rev. D}
  {\bfseries 104} no.~2, (2021) 023521},
  \href{http://arxiv.org/abs/2104.04550}{{\ttfamily arXiv:2104.04550
  [astro-ph.CO]}}.

\bibitem{Barman:2021lot}
B.~Barman and A.~Ghoshal, ``{Scale invariant FIMP miracle},''
  \href{http://dx.doi.org/10.1088/1475-7516/2022/03/003}{{\em JCAP} {\bfseries
  03} no.~03, (2022) 003}, \href{http://arxiv.org/abs/2109.03259}{{\ttfamily
  arXiv:2109.03259 [hep-ph]}}.

\bibitem{Barman:2203}
B.~Barman and A.~Ghoshal, ``{Probing pre-BBN era with Scale Invariant FIMP},''
  \href{http://arxiv.org/abs/2203.13269}{{\ttfamily arXiv:2203.13269}}.

\bibitem{Carone:2013}
C.~D. Carone and R.~Ramos, ``{Classical scale-invariance, the electroweak scale
  and vector dark matter},''
  \href{http://dx.doi.org/10.1103/PhysRevD.88.055020}{{\em Phys. Rev.}
  {\bfseries D88} (2013) 055020},
\href{http://arxiv.org/abs/1307.8428}{{\ttfamily arXiv:1307.8428 [hep-ph]}}.
%%CITATION = ARXIV:1307.8428;%%.

\bibitem{Pelaggi:2014wba}
G.~M. Pelaggi, ``{Predictions of a model of weak scale from dynamical breaking
  of scale invariance},''
  \href{http://dx.doi.org/10.1016/j.nuclphysb.2015.01.025}{{\em Nucl. Phys. B}
  {\bfseries 893} (2015) 443--458},
  \href{http://arxiv.org/abs/1406.4104}{{\ttfamily arXiv:1406.4104 [hep-ph]}}.

\bibitem{Karam:2015}
A.~Karam and K.~Tamvakis, ``{Dark matter and neutrino masses from a
  scale-invariant multi-Higgs portal},''
  \href{http://dx.doi.org/10.1103/PhysRevD.92.075010}{{\em Phys. Rev.}
  {\bfseries D92} no.~7, (2015) 075010},
\href{http://arxiv.org/abs/1508.03031}{{\ttfamily arXiv:1508.03031 [hep-ph]}}.
%%CITATION = ARXIV:1508.03031;%%.

\bibitem{Plascencia:2016}
V.~V. Khoze and A.~D. Plascencia, ``{Dark Matter and Leptogenesis Linked by
  Classical Scale Invariance},''
  \href{http://dx.doi.org/10.1007/JHEP11(2016)025}{{\em JHEP} {\bfseries 11}
  (2016) 025},
\href{http://arxiv.org/abs/1605.06834}{{\ttfamily arXiv:1605.06834 [hep-ph]}}.
%%CITATION = ARXIV:1605.06834;%%.

\bibitem{Chataignier:2018}
L.~Chataignier, T.~Prokopec, M.~G. Schmidt, and B.~\'{S}wie\.{z}ewska,
  ``{Single-scale Renormalisation Group Improvement of Multi-scale Effective
  Potentials},'' \href{http://dx.doi.org/10.1007/JHEP03(2018)014}{{\em JHEP}
  {\bfseries 03} (2018) 014},
\href{http://arxiv.org/abs/1801.05258}{{\ttfamily arXiv:1801.05258 [hep-ph]}}.
%%CITATION = ARXIV:1801.05258;%%.

\bibitem{Marfatia:2020}
D.~Marfatia and P.-Y. Tseng, ``{Gravitational wave signals of dark matter
  freeze-out},'' \href{http://dx.doi.org/10.1007/JHEP02(2021)022}{{\em JHEP}
  {\bfseries 02} (2021) 022}, \href{http://arxiv.org/abs/2006.07313}{{\ttfamily
  arXiv:2006.07313 [hep-ph]}}.

\bibitem{Meissner:2006}
K.~A. Meissner and H.~Nicolai, ``{Conformal Symmetry and the Standard Model},''
  \href{http://dx.doi.org/10.1016/j.physletb.2007.03.023}{{\em Phys. Lett.}
  {\bfseries B648} (2007) 312--317},
\href{http://arxiv.org/abs/hep-th/0612165}{{\ttfamily arXiv:hep-th/0612165
  [hep-th]}}.
%%CITATION = HEP-TH/0612165;%%.

\bibitem{Foot:2007s}
R.~Foot, A.~Kobakhidze, and R.~R. Volkas, ``{Electroweak Higgs as a
  pseudo-Goldstone boson of broken scale invariance},''
  \href{http://dx.doi.org/10.1016/j.physletb.2007.06.084}{{\em Phys. Lett.}
  {\bfseries B655} (2007) 156--161},
\href{http://arxiv.org/abs/0704.1165}{{\ttfamily arXiv:0704.1165 [hep-ph]}}.
%%CITATION = ARXIV:0704.1165;%%.

\bibitem{Foot:2007-3}
R.~Foot, A.~Kobakhidze, K.~McDonald, and R.~Volkas, ``{Neutrino mass in
  radiatively-broken scale-invariant models},''
  \href{http://dx.doi.org/10.1103/PhysRevD.76.075014}{{\em Phys. Rev.}
  {\bfseries D76} (2007) 075014},
\href{http://arxiv.org/abs/0706.1829}{{\ttfamily arXiv:0706.1829 [hep-ph]}}.
%%CITATION = ARXIV:0706.1829;%%.

\bibitem{Foot:2007}
R.~Foot, A.~Kobakhidze, K.~L. McDonald, and R.~R. Volkas, ``{A Solution to the
  hierarchy problem from an almost decoupled hidden sector within a classically
  scale invariant theory},''
  \href{http://dx.doi.org/10.1103/PhysRevD.77.035006}{{\em Phys. Rev.}
  {\bfseries D77} (2008) 035006},
\href{http://arxiv.org/abs/0709.2750}{{\ttfamily arXiv:0709.2750 [hep-ph]}}.
%%CITATION = ARXIV:0709.2750;%%.

\bibitem{Foot:2010av}
R.~Foot, A.~Kobakhidze, and R.~R. Volkas, ``{Stable mass hierarchies and dark
  matter from hidden sectors in the scale-invariant standard model},''
  \href{http://dx.doi.org/10.1103/PhysRevD.82.035005}{{\em Phys. Rev. D}
  {\bfseries 82} (2010) 035005},
  \href{http://arxiv.org/abs/1006.0131}{{\ttfamily arXiv:1006.0131 [hep-ph]}}.

\bibitem{AlexanderNunneley:2010}
L.~Alexander-Nunneley and A.~Pilaftsis, ``{The Minimal Scale Invariant
  Extension of the Standard Model},''
  \href{http://dx.doi.org/10.1007/JHEP09(2010)021}{{\em JHEP} {\bfseries 09}
  (2010) 021},
\href{http://arxiv.org/abs/1006.5916}{{\ttfamily arXiv:1006.5916 [hep-ph]}}.
%%CITATION = ARXIV:1006.5916;%%.

\bibitem{Foot:2010et}
R.~Foot, A.~Kobakhidze, and R.~R. Volkas, ``{Cosmological constant in
  scale-invariant theories},''
  \href{http://dx.doi.org/10.1103/PhysRevD.84.075010}{{\em Phys. Rev. D}
  {\bfseries 84} (2011) 075010},
  \href{http://arxiv.org/abs/1012.4848}{{\ttfamily arXiv:1012.4848 [hep-ph]}}.

\bibitem{Lee:2012}
J.~S. Lee and A.~Pilaftsis, ``{Radiative Corrections to Scalar Masses and
  Mixing in a Scale Invariant Two Higgs Doublet Model},''
  \href{http://dx.doi.org/10.1103/PhysRevD.86.035004}{{\em Phys. Rev.}
  {\bfseries D86} (2012) 035004},
\href{http://arxiv.org/abs/1201.4891}{{\ttfamily arXiv:1201.4891 [hep-ph]}}.
%%CITATION = ARXIV:1201.4891;%%.

\bibitem{Farzinnia:2013}
A.~Farzinnia, H.-J. He, and J.~Ren, ``{Natural Electroweak Symmetry Breaking
  from Scale Invariant Higgs Mechanism},''
  \href{http://dx.doi.org/10.1016/j.physletb.2013.09.060}{{\em Phys. Lett.}
  {\bfseries B727} (2013) 141--150},
\href{http://arxiv.org/abs/1308.0295}{{\ttfamily arXiv:1308.0295 [hep-ph]}}.
%%CITATION = ARXIV:1308.0295;%%.

\bibitem{Gabrielli:2013}
E.~Gabrielli, M.~Heikinheimo, K.~Kannike, A.~Racioppi, M.~Raidal, and
  C.~Spethmann, ``{Towards Completing the Standard Model: Vacuum Stability,
  EWSB and Dark Matter},''
  \href{http://dx.doi.org/10.1103/PhysRevD.89.015017}{{\em Phys. Rev.}
  {\bfseries D89} no.~1, (2014) 015017},
\href{http://arxiv.org/abs/1309.6632}{{\ttfamily arXiv:1309.6632 [hep-ph]}}.
%%CITATION = ARXIV:1309.6632;%%.

\bibitem{Steele:2013fka}
T.~G. Steele, Z.-W. Wang, D.~Contreras, and R.~B. Mann, ``{Viable dark matter
  via radiative symmetry breaking in a scalar singlet Higgs portal extension of
  the standard model},''
  \href{http://dx.doi.org/10.1103/PhysRevLett.112.171602}{{\em Phys. Rev.
  Lett.} {\bfseries 112} no.~17, (2014) 171602},
  \href{http://arxiv.org/abs/1310.1960}{{\ttfamily arXiv:1310.1960 [hep-ph]}}.

\bibitem{Guo:2014}
J.~Guo and Z.~Kang, ``{Higgs Naturalness and Dark Matter Stability by Scale
  Invariance},'' \href{http://dx.doi.org/10.1016/j.nuclphysb.2015.07.014}{{\em
  Nucl. Phys.} {\bfseries B898} (2015) 415--430},
\href{http://arxiv.org/abs/1401.5609}{{\ttfamily arXiv:1401.5609 [hep-ph]}}.
%%CITATION = ARXIV:1401.5609;%%.

\bibitem{Salvio:2014soa}
A.~Salvio and A.~Strumia, ``{Agravity},''
  \href{http://dx.doi.org/10.1007/JHEP06(2014)080}{{\em JHEP} {\bfseries 06}
  (2014) 080}, \href{http://arxiv.org/abs/1403.4226}{{\ttfamily arXiv:1403.4226
  [hep-ph]}}.

\bibitem{Davoudiasl:2014}
H.~Davoudiasl and I.~M. Lewis, ``{Right-Handed Neutrinos as the Origin of the
  Electroweak Scale},''
  \href{http://dx.doi.org/10.1103/PhysRevD.90.033003}{{\em Phys. Rev.}
  {\bfseries D90} no.~3, (2014) 033003},
\href{http://arxiv.org/abs/1404.6260}{{\ttfamily arXiv:1404.6260 [hep-ph]}}.
%%CITATION = ARXIV:1404.6260;%%.

\bibitem{Farzinnia:2014xia}
A.~Farzinnia and J.~Ren, ``{Higgs Partner Searches and Dark Matter
  Phenomenology in a Classically Scale Invariant Higgs Boson Sector},''
  \href{http://dx.doi.org/10.1103/PhysRevD.90.015019}{{\em Phys. Rev. D}
  {\bfseries 90} no.~1, (2014) 015019},
  \href{http://arxiv.org/abs/1405.0498}{{\ttfamily arXiv:1405.0498 [hep-ph]}}.

\bibitem{Lindner:2014oea}
M.~Lindner, S.~Schmidt, and J.~Smirnov, ``{Neutrino Masses and Conformal
  Electro-Weak Symmetry Breaking},''
  \href{http://dx.doi.org/10.1007/JHEP10(2014)177}{{\em JHEP} {\bfseries 10}
  (2014) 177}, \href{http://arxiv.org/abs/1405.6204}{{\ttfamily arXiv:1405.6204
  [hep-ph]}}.

\bibitem{Kang:2014cia}
Z.~Kang, ``{Upgrading sterile neutrino dark matter to FI$m$P using scale
  invariance},'' \href{http://dx.doi.org/10.1140/epjc/s10052-015-3702-4}{{\em
  Eur. Phys. J. C} {\bfseries 75} no.~10, (2015) 471},
  \href{http://arxiv.org/abs/1411.2773}{{\ttfamily arXiv:1411.2773 [hep-ph]}}.

\bibitem{Kannike:2015apa}
K.~Kannike, G.~H\"utsi, L.~Pizza, A.~Racioppi, M.~Raidal, A.~Salvio, and
  A.~Strumia, ``{Dynamically Induced Planck Scale and Inflation},''
  \href{http://dx.doi.org/10.1007/JHEP05(2015)065}{{\em JHEP} {\bfseries 05}
  (2015) 065}, \href{http://arxiv.org/abs/1502.01334}{{\ttfamily
  arXiv:1502.01334 [astro-ph.CO]}}.

\bibitem{Endo:2015ifa}
K.~Endo and Y.~Sumino, ``{A Scale-invariant Higgs Sector and Structure of the
  Vacuum},'' \href{http://dx.doi.org/10.1007/JHEP05(2015)030}{{\em JHEP}
  {\bfseries 05} (2015) 030}, \href{http://arxiv.org/abs/1503.02819}{{\ttfamily
  arXiv:1503.02819 [hep-ph]}}.

\bibitem{Kang:2015aqa}
Z.~Kang, ``{View FImP miracle (by scale invariance) \`a la self-interaction},''
  \href{http://dx.doi.org/10.1016/j.physletb.2015.10.031}{{\em Phys. Lett. B}
  {\bfseries 751} (2015) 201--204},
  \href{http://arxiv.org/abs/1505.06554}{{\ttfamily arXiv:1505.06554
  [hep-ph]}}.

\bibitem{Endo:2015nba}
K.~Endo and K.~Ishiwata, ``{Direct detection of singlet dark matter in
  classically scale-invariant standard model},''
  \href{http://dx.doi.org/10.1016/j.physletb.2015.08.059}{{\em Phys. Lett. B}
  {\bfseries 749} (2015) 583--588},
  \href{http://arxiv.org/abs/1507.01739}{{\ttfamily arXiv:1507.01739
  [hep-ph]}}.

\bibitem{Ahriche:2015loa}
A.~Ahriche, K.~L. McDonald, and S.~Nasri, ``{A Radiative Model for the Weak
  Scale and Neutrino Mass via Dark Matter},''
  \href{http://dx.doi.org/10.1007/JHEP02(2016)038}{{\em JHEP} {\bfseries 02}
  (2016) 038}, \href{http://arxiv.org/abs/1508.02607}{{\ttfamily
  arXiv:1508.02607 [hep-ph]}}.

\bibitem{Wang:2015cda}
Z.-W. Wang, T.~G. Steele, T.~Hanif, and R.~B. Mann, ``{Conformal Complex
  Singlet Extension of the Standard Model: Scenario for Dark Matter and a
  Second Higgs Boson},'' \href{http://dx.doi.org/10.1007/JHEP08(2016)065}{{\em
  JHEP} {\bfseries 08} (2016) 065},
  \href{http://arxiv.org/abs/1510.04321}{{\ttfamily arXiv:1510.04321
  [hep-ph]}}.

\bibitem{Ghorbani:2015xvz}
K.~Ghorbani and H.~Ghorbani, ``{Scalar Dark Matter in Scale Invariant Standard
  Model},'' \href{http://dx.doi.org/10.1007/JHEP04(2016)024}{{\em JHEP}
  {\bfseries 04} (2016) 024}, \href{http://arxiv.org/abs/1511.08432}{{\ttfamily
  arXiv:1511.08432 [hep-ph]}}.

\bibitem{Farzinnia:2015fka}
A.~Farzinnia and S.~Kouwn, ``{Classically scale invariant inflation,
  supermassive WIMPs, and adimensional gravity},''
  \href{http://dx.doi.org/10.1103/PhysRevD.93.063528}{{\em Phys. Rev. D}
  {\bfseries 93} no.~6, (2016) 063528},
  \href{http://arxiv.org/abs/1512.05890}{{\ttfamily arXiv:1512.05890
  [hep-ph]}}.

\bibitem{Helmboldt:2016}
A.~J. Helmboldt, P.~Humbert, M.~Lindner, and J.~Smirnov, ``{Minimal conformal
  extensions of the Higgs sector},''
  \href{http://dx.doi.org/10.1007/JHEP07(2017)113}{{\em JHEP} {\bfseries 07}
  (2017) 113},
\href{http://arxiv.org/abs/1603.03603}{{\ttfamily arXiv:1603.03603 [hep-ph]}}.
%%CITATION = ARXIV:1603.03603;%%.

\bibitem{Ahriche:2016cio}
A.~Ahriche, K.~L. McDonald, and S.~Nasri, ``{The Scale-Invariant Scotogenic
  Model},'' \href{http://dx.doi.org/10.1007/JHEP06(2016)182}{{\em JHEP}
  {\bfseries 06} (2016) 182}, \href{http://arxiv.org/abs/1604.05569}{{\ttfamily
  arXiv:1604.05569 [hep-ph]}}.

\bibitem{Ahriche:2016ixu}
A.~Ahriche, A.~Manning, K.~L. McDonald, and S.~Nasri, ``{Scale-Invariant Models
  with One-Loop Neutrino Mass and Dark Matter Candidates},''
  \href{http://dx.doi.org/10.1103/PhysRevD.94.053005}{{\em Phys. Rev. D}
  {\bfseries 94} no.~5, (2016) 053005},
  \href{http://arxiv.org/abs/1604.05995}{{\ttfamily arXiv:1604.05995
  [hep-ph]}}.

\bibitem{Wu:2016jdo}
F.~Wu, ``{Aspects of a nonminimal conformal extension of the standard model},''
  \href{http://dx.doi.org/10.1103/PhysRevD.94.055011}{{\em Phys. Rev. D}
  {\bfseries 94} no.~5, (2016) 055011},
  \href{http://arxiv.org/abs/1606.08112}{{\ttfamily arXiv:1606.08112
  [hep-ph]}}.

\bibitem{YaserAyazi:2018lrv}
S.~Yaser~Ayazi and A.~Mohamadnejad, ``{Scale-Invariant Two Component Dark
  Matter},'' \href{http://dx.doi.org/10.1140/epjc/s10052-019-6651-5}{{\em Eur.
  Phys. J. C} {\bfseries 79} no.~2, (2019) 140},
  \href{http://arxiv.org/abs/1808.08706}{{\ttfamily arXiv:1808.08706
  [hep-ph]}}.

\bibitem{Oda:2018zth}
I.~Oda, ``{Planck and Electroweak Scales Emerging from Conformal Gravity},''
  \href{http://dx.doi.org/10.1140/epjc/s10052-018-6289-8}{{\em Eur. Phys. J. C}
  {\bfseries 78} no.~10, (2018) 798},
  \href{http://arxiv.org/abs/1806.03420}{{\ttfamily arXiv:1806.03420
  [hep-th]}}.

\bibitem{Brdar:2018}
V.~Brdar, Y.~Emonds, A.~J. Helmboldt, and M.~Lindner, ``{Conformal Realization
  of the Neutrino Option},''
  \href{http://dx.doi.org/10.1103/PhysRevD.99.055014}{{\em Phys. Rev. D}
  {\bfseries 99} no.~5, (2019) 055014},
  \href{http://arxiv.org/abs/1807.11490}{{\ttfamily arXiv:1807.11490
  [hep-ph]}}.

\bibitem{Brdar:2018num}
V.~Brdar, A.~J. Helmboldt, and J.~Kubo, ``{Gravitational Waves from First-Order
  Phase Transitions: LIGO as a Window to Unexplored Seesaw Scales},''
  \href{http://dx.doi.org/10.1088/1475-7516/2019/02/021}{{\em JCAP} {\bfseries
  02} (2019) 021}, \href{http://arxiv.org/abs/1810.12306}{{\ttfamily
  arXiv:1810.12306 [hep-ph]}}.

\bibitem{Mohamadnejad:2019wqb}
A.~Mohamadnejad, ``{Accidental scale-invariant Majorana dark matter in
  leptoquark-Higgs portals},''
  \href{http://dx.doi.org/10.1016/j.nuclphysb.2019.114793}{{\em Nucl. Phys. B}
  {\bfseries 949} (2019) 114793},
  \href{http://arxiv.org/abs/1904.03857}{{\ttfamily arXiv:1904.03857
  [hep-ph]}}.

\bibitem{Kannike:2019upf}
K.~Kannike, A.~Kubarski, and L.~Marzola, ``{Geometry of Flat Directions in
  Scale-Invariant Potentials},''
  \href{http://dx.doi.org/10.1103/PhysRevD.99.115034}{{\em Phys. Rev. D}
  {\bfseries 99} no.~11, (2019) 115034},
  \href{http://arxiv.org/abs/1904.07867}{{\ttfamily arXiv:1904.07867
  [hep-ph]}}.

\bibitem{Jung:2019dog}
D.-W. Jung, J.~Lee, and S.-H. Nam, ``{Scalar dark matter in the conformally
  invariant extension of the standard model},''
  \href{http://dx.doi.org/10.1016/j.physletb.2019.134823}{{\em Phys. Lett. B}
  {\bfseries 797} (2019) 134823},
  \href{http://arxiv.org/abs/1904.10209}{{\ttfamily arXiv:1904.10209
  [hep-ph]}}.

\bibitem{Brdar:2019qut}
V.~Brdar, A.~J. Helmboldt, and M.~Lindner, ``{Strong Supercooling as a
  Consequence of Renormalization Group Consistency},''
  \href{http://dx.doi.org/10.1007/JHEP12(2019)158}{{\em JHEP} {\bfseries 12}
  (2019) 158}, \href{http://arxiv.org/abs/1910.13460}{{\ttfamily
  arXiv:1910.13460 [hep-ph]}}.

\bibitem{Braathen:2020vwo}
J.~Braathen, S.~Kanemura, and M.~Shimoda, ``{Two-loop analysis of classically
  scale-invariant models with extended Higgs sectors},''
  \href{http://dx.doi.org/10.1007/JHEP03(2021)297}{{\em JHEP} {\bfseries 03}
  (2021) 297}, \href{http://arxiv.org/abs/2011.07580}{{\ttfamily
  arXiv:2011.07580 [hep-ph]}}.

\bibitem{Kannike:2020ppf}
K.~Kannike, K.~Loos, and L.~Marzola, ``{Minima of classically scale-invariant
  potentials},'' \href{http://dx.doi.org/10.1007/JHEP06(2021)128}{{\em JHEP}
  {\bfseries 06} (2021) 128}, \href{http://arxiv.org/abs/2011.12304}{{\ttfamily
  arXiv:2011.12304 [hep-ph]}}.

\bibitem{Kubo:2020fdd}
J.~Kubo, J.~Kuntz, M.~Lindner, J.~Rezacek, P.~Saake, and A.~Trautner,
  ``{Unified emergence of energy scales and cosmic inflation},''
  \href{http://dx.doi.org/10.1007/JHEP08(2021)016}{{\em JHEP} {\bfseries 08}
  (2021) 016}, \href{http://arxiv.org/abs/2012.09706}{{\ttfamily
  arXiv:2012.09706 [hep-ph]}}.

\bibitem{Ahriche:2021frb}
A.~Ahriche, ``{Purely Radiative Higgs Mass in Scale invariant models},''
  \href{http://dx.doi.org/10.1016/j.nuclphysb.2022.115896}{{\em Nucl. Phys. B}
  {\bfseries 982} (2022) 115896},
  \href{http://arxiv.org/abs/2110.10301}{{\ttfamily arXiv:2110.10301
  [hep-ph]}}.

\bibitem{Soualah:2021xbn}
R.~Soualah and A.~Ahriche, ``{Scale invariant scotogenic model: Dark matter and
  the scalar sector},''
  \href{http://dx.doi.org/10.1103/PhysRevD.105.055017}{{\em Phys. Rev. D}
  {\bfseries 105} no.~5, (2022) 055017},
  \href{http://arxiv.org/abs/2111.01121}{{\ttfamily arXiv:2111.01121
  [hep-ph]}}.

\bibitem{Dias:2005jk}
A.~G. Dias, C.~A. de~S.~Pires, V.~Pleitez, and P.~S. Rodrigues~da Silva,
  ``{Dynamically induced spontaneous symmetry breaking in 3-3-1 models},''
  \href{http://dx.doi.org/10.1016/j.physletb.2005.06.048}{{\em Phys. Lett. B}
  {\bfseries 621} (2005) 151--159},
  \href{http://arxiv.org/abs/hep-ph/0503192}{{\ttfamily arXiv:hep-ph/0503192}}.

\bibitem{Holthausen:2009uc}
M.~Holthausen, M.~Lindner, and M.~A. Schmidt, ``{Radiative Symmetry Breaking of
  the Minimal Left-Right Symmetric Model},''
  \href{http://dx.doi.org/10.1103/PhysRevD.82.055002}{{\em Phys. Rev. D}
  {\bfseries 82} (2010) 055002},
  \href{http://arxiv.org/abs/0911.0710}{{\ttfamily arXiv:0911.0710 [hep-ph]}}.

\bibitem{Heikinheimo:2013}
M.~Heikinheimo, A.~Racioppi, M.~Raidal, C.~Spethmann, and K.~Tuominen,
  ``{Physical Naturalness and Dynamical Breaking of Classical Scale
  Invariance},'' \href{http://dx.doi.org/10.1142/S0217732314500771}{{\em Mod.
  Phys. Lett.} {\bfseries A29} (2014) 1450077},
\href{http://arxiv.org/abs/1304.7006}{{\ttfamily arXiv:1304.7006 [hep-ph]}}.
%%CITATION = ARXIV:1304.7006;%%.

\bibitem{Dermisek:2013}
D.~Chway, T.~H. Jung, H.~D. Kim, and R.~Dermisek, ``{Radiative Electroweak
  Symmetry Breaking Model Perturbative All the Way to the Planck Scale},''
  \href{http://dx.doi.org/10.1103/PhysRevLett.113.051801}{{\em Phys. Rev.
  Lett.} {\bfseries 113} no.~5, (2014) 051801},
\href{http://arxiv.org/abs/1308.0891}{{\ttfamily arXiv:1308.0891 [hep-ph]}}.
%%CITATION = ARXIV:1308.0891;%%.

\bibitem{Holthausen:2013ota}
M.~Holthausen, J.~Kubo, K.~S. Lim, and M.~Lindner, ``{Electroweak and Conformal
  Symmetry Breaking by a Strongly Coupled Hidden Sector},''
  \href{http://dx.doi.org/10.1007/JHEP12(2013)076}{{\em JHEP} {\bfseries 12}
  (2013) 076}, \href{http://arxiv.org/abs/1310.4423}{{\ttfamily arXiv:1310.4423
  [hep-ph]}}.

\bibitem{Kubo:2014ida}
J.~Kubo, K.~S. Lim, and M.~Lindner, ``{Gamma-ray Line from Nambu-Goldstone Dark
  Matter in a Scale Invariant Extension of the Standard Model},''
  \href{http://dx.doi.org/10.1007/JHEP09(2014)016}{{\em JHEP} {\bfseries 09}
  (2014) 016}, \href{http://arxiv.org/abs/1405.1052}{{\ttfamily arXiv:1405.1052
  [hep-ph]}}.

\bibitem{Altmannshofer:2014}
W.~Altmannshofer, W.~A. Bardeen, M.~Bauer, M.~Carena, and J.~D. Lykken,
  ``{Light Dark Matter, Naturalness, and the Radiative Origin of the
  Electroweak Scale},'' \href{http://dx.doi.org/10.1007/JHEP01(2015)032}{{\em
  JHEP} {\bfseries 01} (2015) 032},
\href{http://arxiv.org/abs/1408.3429}{{\ttfamily arXiv:1408.3429 [hep-ph]}}.
%%CITATION = ARXIV:1408.3429;%%.

\bibitem{Antipin:2014qva}
O.~Antipin, M.~Redi, and A.~Strumia, ``{Dynamical generation of the weak and
  Dark Matter scales from strong interactions},''
  \href{http://dx.doi.org/10.1007/JHEP01(2015)157}{{\em JHEP} {\bfseries 01}
  (2015) 157}, \href{http://arxiv.org/abs/1410.1817}{{\ttfamily arXiv:1410.1817
  [hep-ph]}}.

\bibitem{Giudice:2014tma}
G.~F. Giudice, G.~Isidori, A.~Salvio, and A.~Strumia, ``{Softened Gravity and
  the Extension of the Standard Model up to Infinite Energy},''
  \href{http://dx.doi.org/10.1007/JHEP02(2015)137}{{\em JHEP} {\bfseries 02}
  (2015) 137}, \href{http://arxiv.org/abs/1412.2769}{{\ttfamily arXiv:1412.2769
  [hep-ph]}}.

\bibitem{Ametani:2015jla}
Y.~Ametani, M.~Aoki, H.~Goto, and J.~Kubo, ``{Nambu-Goldstone Dark Matter in a
  Scale Invariant Bright Hidden Sector},''
  \href{http://dx.doi.org/10.1103/PhysRevD.91.115007}{{\em Phys. Rev. D}
  {\bfseries 91} no.~11, (2015) 115007},
  \href{http://arxiv.org/abs/1505.00128}{{\ttfamily arXiv:1505.00128
  [hep-ph]}}.

\bibitem{Carone:2015jra}
C.~D. Carone and R.~Ramos, ``{Dark chiral symmetry breaking and the origin of
  the electroweak scale},''
  \href{http://dx.doi.org/10.1016/j.physletb.2015.05.044}{{\em Phys. Lett. B}
  {\bfseries 746} (2015) 424--429},
  \href{http://arxiv.org/abs/1505.04448}{{\ttfamily arXiv:1505.04448
  [hep-ph]}}.

\bibitem{Kubo:2015joa}
J.~Kubo and M.~Yamada, ``{Scale and electroweak first-order phase
  transitions},'' \href{http://dx.doi.org/10.1093/ptep/ptv114}{{\em PTEP}
  {\bfseries 2015} no.~9, (2015) 093B01},
  \href{http://arxiv.org/abs/1506.06460}{{\ttfamily arXiv:1506.06460
  [hep-ph]}}.

\bibitem{Latosinski:2015pba}
A.~Latosinski, A.~Lewandowski, K.~A. Meissner, and H.~Nicolai, ``{Conformal
  Standard Model with an extended scalar sector},''
  \href{http://dx.doi.org/10.1007/JHEP10(2015)170}{{\em JHEP} {\bfseries 10}
  (2015) 170}, \href{http://arxiv.org/abs/1507.01755}{{\ttfamily
  arXiv:1507.01755 [hep-ph]}}.

\bibitem{Haba:2015qbz}
N.~Haba, H.~Ishida, N.~Kitazawa, and Y.~Yamaguchi, ``{A new dynamics of
  electroweak symmetry breaking with classically scale invariance},''
  \href{http://dx.doi.org/10.1016/j.physletb.2016.02.052}{{\em Phys. Lett. B}
  {\bfseries 755} (2016) 439--443},
  \href{http://arxiv.org/abs/1512.05061}{{\ttfamily arXiv:1512.05061
  [hep-ph]}}.

\bibitem{Karam:2016}
A.~Karam and K.~Tamvakis, ``{Dark Matter from a Classically Scale-Invariant
  $SU(3)_X$},'' \href{http://dx.doi.org/10.1103/PhysRevD.94.055004}{{\em Phys.
  Rev.} {\bfseries D94} no.~5, (2016) 055004},
\href{http://arxiv.org/abs/1607.01001}{{\ttfamily arXiv:1607.01001 [hep-ph]}}.
%%CITATION = ARXIV:1607.01001;%%.

\bibitem{Kubo:2016}
J.~Kubo and M.~Yamada, ``{Scale genesis and gravitational wave in a classically
  scale invariant extension of the standard model},''
  \href{http://dx.doi.org/10.1088/1475-7516/2016/12/001}{{\em JCAP} {\bfseries
  1612} no.~12, (2016) 001},
\href{http://arxiv.org/abs/1610.02241}{{\ttfamily arXiv:1610.02241 [hep-ph]}}.
%%CITATION = ARXIV:1610.02241;%%.

\bibitem{Ishida:2019gri}
H.~Ishida, S.~Matsuzaki, and R.~Ouyang, ``{Unified interpretation of
  scalegenesis in conformally extended standard models: a dynamical origin of
  Higgs portal},'' \href{http://dx.doi.org/10.1088/1674-1137/abb07f}{{\em Chin.
  Phys. C} {\bfseries 44} no.~11, (2020) 111002},
  \href{http://arxiv.org/abs/1907.09176}{{\ttfamily arXiv:1907.09176
  [hep-ph]}}.

\bibitem{Dias:2020ryz}
A.~G. Dias, J.~Leite, B.~L. S\'anchez-Vega, and W.~C. Vieira, ``{Dynamical
  symmetry breaking and fermion mass hierarchy in the scale-invariant 3-3-1
  model},'' \href{http://dx.doi.org/10.1103/PhysRevD.102.015021}{{\em Phys.
  Rev. D} {\bfseries 102} no.~1, (2020) 015021},
  \href{http://arxiv.org/abs/2005.00556}{{\ttfamily arXiv:2005.00556
  [hep-ph]}}.

\bibitem{Aoki:2020mlo}
M.~Aoki, V.~Brdar, and J.~Kubo, ``{Heavy dark matter, neutrino masses, and
  Higgs naturalness from a strongly interacting hidden sector},''
  \href{http://dx.doi.org/10.1103/PhysRevD.102.035026}{{\em Phys. Rev. D}
  {\bfseries 102} no.~3, (2020) 035026},
  \href{http://arxiv.org/abs/2007.04367}{{\ttfamily arXiv:2007.04367
  [hep-ph]}}.

\bibitem{Dias:2022hbu}
A.~G. Dias, J.~Leite, and B.~L. S\'anchez-Vega, ``{Scale-invariant 3-3-1-1
  model with $B-L$ symmetry},''
  \href{http://arxiv.org/abs/2207.06276}{{\ttfamily arXiv:2207.06276
  [hep-ph]}}.

\bibitem{Gross:2015}
C.~Gross, O.~Lebedev, and Y.~Mambrini, ``{Non-Abelian gauge fields as dark
  matter},'' \href{http://dx.doi.org/10.1007/JHEP08(2015)158}{{\em JHEP}
  {\bfseries 08} (2015) 158},
\href{http://arxiv.org/abs/1505.07480}{{\ttfamily arXiv:1505.07480 [hep-ph]}}.
%%CITATION = ARXIV:1505.07480;%%.

\bibitem{Hambye:2008}
T.~Hambye, ``{Hidden vector dark matter},''
  \href{http://dx.doi.org/10.1088/1126-6708/2009/01/028}{{\em JHEP} {\bfseries
  01} (2009) 028},
\href{http://arxiv.org/abs/0811.0172}{{\ttfamily arXiv:0811.0172 [hep-ph]}}.
%%CITATION = ARXIV:0811.0172;%%.

\bibitem{Croon:2020}
D.~Croon, O.~Gould, P.~Schicho, T.~V.~I. Tenkanen, and G.~White, ``{Theoretical
  uncertainties for cosmological first-order phase transitions},''
  \href{http://dx.doi.org/10.1007/JHEP04(2021)055}{{\em JHEP} {\bfseries 04}
  (2021) 055}, \href{http://arxiv.org/abs/2009.10080}{{\ttfamily
  arXiv:2009.10080 [hep-ph]}}.

\bibitem{Athron:2022}
P.~Athron, C.~Balazs, A.~Fowlie, L.~Morris, G.~White, and Y.~Zhang, ``{How
  arbitrary are perturbative calculations of the electroweak phase
  transition?},'' \href{http://arxiv.org/abs/2208.01319}{{\ttfamily
  arXiv:2208.01319 [hep-ph]}}.

\bibitem{Ellis:2018}
J.~Ellis, M.~Lewicki, and J.~M. No, ``{On the Maximal Strength of a First-Order
  Electroweak Phase Transition and its Gravitational Wave Signal},''
  \href{http://dx.doi.org/10.1088/1475-7516/2019/04/003}{{\em JCAP} {\bfseries
  04} (2019) 003}, \href{http://arxiv.org/abs/1809.08242}{{\ttfamily
  arXiv:1809.08242 [hep-ph]}}.

\bibitem{Ellis:2019}
J.~Ellis, M.~Lewicki, J.~M. No, and V.~Vaskonen, ``{Gravitational wave energy
  budget in strongly supercooled phase transitions},''
  \href{http://dx.doi.org/10.1088/1475-7516/2019/06/024}{{\em JCAP} {\bfseries
  06} (2019) 024}, \href{http://arxiv.org/abs/1903.09642}{{\ttfamily
  arXiv:1903.09642 [hep-ph]}}.

\bibitem{Lewicki:2019}
M.~Lewicki and V.~Vaskonen, ``{On bubble collisions in strongly supercooled
  phase transitions},''
  \href{http://dx.doi.org/10.1016/j.dark.2020.100672}{{\em Phys. Dark Univ.}
  {\bfseries 30} (2020) 100672},
  \href{http://arxiv.org/abs/1912.00997}{{\ttfamily arXiv:1912.00997
  [astro-ph.CO]}}.

\bibitem{Ellis:2020-2}
J.~Ellis, M.~Lewicki, and J.~M. No, ``{Gravitational waves from first-order
  cosmological phase transitions: lifetime of the sound wave source},''
  \href{http://dx.doi.org/10.1088/1475-7516/2020/07/050}{{\em JCAP} {\bfseries
  07} (2020) 050}, \href{http://arxiv.org/abs/2003.07360}{{\ttfamily
  arXiv:2003.07360 [hep-ph]}}.

\bibitem{Ellis:2020}
J.~Ellis, M.~Lewicki, and V.~Vaskonen, ``{Updated predictions for gravitational
  waves produced in a strongly supercooled phase transition},''
  \href{http://dx.doi.org/10.1088/1475-7516/2020/11/020}{{\em JCAP} {\bfseries
  11} (2020) 020}, \href{http://arxiv.org/abs/2007.15586}{{\ttfamily
  arXiv:2007.15586 [astro-ph.CO]}}.

\bibitem{XENON:2018voc}
{\bfseries XENON} Collaboration, E.~Aprile {\em et~al.}, ``{Dark Matter Search
  Results from a One Ton-Year Exposure of XENON1T},''
  \href{http://dx.doi.org/10.1103/PhysRevLett.121.111302}{{\em Phys. Rev.
  Lett.} {\bfseries 121} no.~11, (2018) 111302},
  \href{http://arxiv.org/abs/1805.12562}{{\ttfamily arXiv:1805.12562
  [astro-ph.CO]}}.

\bibitem{PandaX-4T:2021bab}
{\bfseries PandaX-4T} Collaboration, Y.~Meng {\em et~al.}, ``{Dark Matter
  Search Results from the PandaX-4T Commissioning Run},''
  \href{http://dx.doi.org/10.1103/PhysRevLett.127.261802}{{\em Phys. Rev.
  Lett.} {\bfseries 127} no.~26, (2021) 261802},
  \href{http://arxiv.org/abs/2107.13438}{{\ttfamily arXiv:2107.13438
  [hep-ex]}}.

\bibitem{LZ:2022ufs}
{\bfseries LZ} Collaboration, J.~Aalbers {\em et~al.}, ``{First Dark Matter
  Search Results from the LUX-ZEPLIN (LZ) Experiment},''
  \href{http://arxiv.org/abs/2207.03764}{{\ttfamily arXiv:2207.03764
  [hep-ex]}}.

\bibitem{XENON:2020kmp}
{\bfseries XENON} Collaboration, E.~Aprile {\em et~al.}, ``{Projected WIMP
  sensitivity of the XENONnT dark matter experiment},''
  \href{http://dx.doi.org/10.1088/1475-7516/2020/11/031}{{\em JCAP} {\bfseries
  11} (2020) 031}, \href{http://arxiv.org/abs/2007.08796}{{\ttfamily
  arXiv:2007.08796 [physics.ins-det]}}.

\bibitem{Frandsen:2022klh}
M.~T. Frandsen, M.~E. Thing, M.~Heikinheimo, K.~Tuominen, and M.~Rosenlyst,
  ``{Vector dark matter in supercooled Higgs portal models},''
  \href{http://arxiv.org/abs/2301.00041}{{\ttfamily arXiv:2301.00041
  [hep-ph]}}.

\bibitem{Caprini:2019egz}
C.~Caprini {\em et~al.}, ``{Detecting gravitational waves from cosmological
  phase transitions with LISA: an update},''
  \href{http://dx.doi.org/10.1088/1475-7516/2020/03/024}{{\em JCAP} {\bfseries
  03} (2020) 024}, \href{http://arxiv.org/abs/1910.13125}{{\ttfamily
  arXiv:1910.13125 [astro-ph.CO]}}.

\bibitem{Robson:2018ifk}
T.~Robson, N.~J. Cornish, and C.~Liu, ``{The construction and use of LISA
  sensitivity curves},'' \href{http://dx.doi.org/10.1088/1361-6382/ab1101}{{\em
  Class. Quant. Grav.} {\bfseries 36} no.~10, (2019) 105011},
  \href{http://arxiv.org/abs/1803.01944}{{\ttfamily arXiv:1803.01944
  [astro-ph.HE]}}.

\end{thebibliography}\endgroup
\bibliographystyle{utphys}

% \begin{thebibliography}{99}
% \bibitem{...}
% ....

% \end{thebibliography}

\end{document}